\documentclass[aps,prd,preprint,tightening,showpacs,nofootinbib]{revtex4-2}
\usepackage{amsmath,amssymb,amsthm,graphicx}
\usepackage[mathscr]{eucal}
\usepackage[colorlinks,linkcolor=blue,anchorcolor=blue,citecolor=green]{hyperref}
\usepackage[dvipsnames,usenames]{color}
\usepackage{graphicx}
\usepackage{verbatim,color,ulem}
\usepackage{epsf,epsfig,graphics}
\usepackage{verbatim,color,ulem}
\usepackage{physics}
\usepackage{subeqnarray}
\usepackage{orcidlink}

\begin{document}
\title{Motion of spinning particles in the Kerr–Newman black hole exterior}
\author{Yi-Ping Chen \orcidlink{0009-0003-2462-2592}}
\author{Tien Hsieh \orcidlink{0000-0001-7199-1241}}
\author{Da-Shin Lee \orcidlink{0000-0003-3187-8863}}
\email{dslee@gms.ndhu.edu.tw}
\affiliation{
Department of Physics, National Dong Hwa University, Hualien, Taiwan, Republic of China}
\date{\today}
\begin{abstract}
The motion of a spinning particle in the exterior of a Kerr–Newman black hole is studied.
The dynamics is governed by the Mathisson–Papapetrou equations in the pole-dipole approximation, which includes spin-curvature coupling to the first order of spin.
In terms of conserved quantities, the dynamical equations in Mino time can be transformed into the integral form for both aligned and misaligned spins with respect to the orbital motion.
These non-geodesic equations can be solved analytically, and the solutions involve Jacobi elliptic functions.
We derive the radial potential to study the parameter space of the particle for various types of orbits based on its roots, corrected by the particle's spin.
In the misaligned case, we consider equatorial motion of a particle oscillating between two turning points, which are the two outermost roots of the radial potential.
This results in an induced oscillatory motion out of the equatorial plane.
In particular, the periods of the motion are obtained explicitly.
To validate our analytical solutions further, we compare them with the results of exact numerical integration, demonstrating good agreement.
When the orbits become a source of gravitational-wave emission, these periods of motion will provide essential input in determining gravitational-wave signals in the frequency domain.
The implications for gravitational-wave emission due to extreme mass-ratio inspirals (EMRIs) are discussed. 
\end{abstract}
\maketitle
\newpage
\section{Introduction}
The discovery of gravitational waves from compact binary inspirals by LIGO has revolutionized gravitational-wave astronomy and opened up a new avenue of research \cite{abbott-2016, abbott-2019, abbott-2021}.
The advent of space-based gravitational-wave detectors such as the planned Laser Interferometer Space Antenna (LISA) is driven by the need to detect gravitational-wave sources in the millihertz frequency band, where one of the key sources is from extreme mass-ratio inspirals (EMRIs) \cite{consortium-2013, barausse-2020, group-2023, babak-2017, kocsis-2011, danzmann-2000, prince-2003, amaro-seoane-2012}.
In astrophysics, EMRIs consist of a stellar mass object such as a black hole or neutron star of mass $ 1-10^2 M_\odot$ orbiting a massive black hole of mass of $ 10^5-10^7 M_\odot$, resulting in a binary with an extreme mass ratio.
The binary gradually loses energy and angular momentum due to the emission of gravitational waves, causing the stellar-mass object to eventually spiral into the massive black hole.

A fairly good approximation of the motion of EMRIs is to consider the trajectory of a particle orbiting a black hole.
The motion of a spinless particle around Kerr family black holes along geodesics has been extensively studied in the literature \cite{gralla-2020, fujita-2009} and by us \cite{wang-2022, li-2023, ko-2024}.
The orbital solutions can be expressed in terms of elliptic integrals and Jacobi elliptic functions in Mino time \cite{w-1965,mino-2003, gralla-2020}.
The closed-form solutions for the motion of a spinning particle near a Schwarzschild black hole and a Kerr black hole are studied in \cite{witzany-2024} and \cite{skoupy-2025A, skoupy-2025B} to the leading order in the spin of a particle.
Theoretical considerations, together with recent observations of structures near Sagittarius A* by the GRAVITY experiment \cite{abuter-2018}, indicate the possible presence of a small electric charge of the central massive black hole \cite{zajacek-2018, zajacek-2019}.
Thus, it is of great interest to explore the motion of the particle around a charged black hole.
In \cite{ciou-2025}, we consider a spinning particle orbiting the exterior of a charged black hole \cite{witzany-2024}.
The radius of the innermost stable circular orbits (ISCOs) involving the particle's spin effect was explored.
Then the closed-form solutions for the motion were obtained for both bound and unbound orbits of a particle with aligned and misaligned spins with respect to the orbital angular momentum.
These analytical solutions can be used as input for future gravitational waveform modeling from EMRIs \cite{group-2023, drummond-2022A, drummond-2022B, cui-2025, barack-2004, saijo-1998} as well as the understanding of the black hole accretion \cite{fabian-2020, page-1974, reynolds-1997, schnittman-2016}.

The current work extends our earlier paper \cite{ciou-2025} by considering a spinning particle oscillating between two turning points on the equatorial plane of the exterior of the Kerr–Newman black hole. 
These turning points are the two outermost roots of the radial potential in the misaligned case.
These solutions are given by the reference trajectory with the corrections from the particle's spin.
The solutions here are split in such a way that allows them to be reduced to the known solutions as in \cite{witzany-2024, ciou-2025}, as well as to those for the motion of a spinless particle in a Kerr–Newman black hole in \cite{wang-2022}.
Another advantage is that the resulting solutions exhibit no apparent singular behavior  at the turning points. This feature can be compared with those in \cite{skoupy-2022, skoupy-2023, drummond-2024, piovano-2025A, piovano-2025B, drummond-2026} via the gauge fixing.
Additionally, the solutions are valid to first order in the spin; however, some higher-order spin dependence is implicitly retained, as in \cite{witzany-2024, ciou-2025}. 
We compare the analytical solutions  with the results of exact numerical integration and demonstrate their agreement.
In \cite{babak-2007}, the kludge method computes the gravitational waveform  by combining the particle's trajectories in Boyer–Lindquist coordinates with the linearized gravitational perturbation equations in flat spacetime.
This method employs the slow-motion quadrupole or quadrupole-octupole formula reduced from a more general formula for fast motion, but in the weak field scenario \cite{press-1977}.
We apply this approach to give an order-of-magnitude estimate of the gravitational-wave emission due to EMRIs.
The \texttt{Mathematica} codes for all calculations are provided by us in \cite{chen_2026_21881833}.
The analytical solutions make the dependence on conserved quantities explicit and allow for the efficient evaluation of orbital frequencies and turning points. 
They also provide a consistent framework for the calculation of orbital solutions using a particular perturbation expansion.
These advantages are particularly significant in EMRI studies because generating waveforms requires repeatedly evaluating orbital trajectories across a broad range of parameters.

The article is organized as follows.
In Sec. \ref{sec2}, the equations of motion for a spinning particle in the exterior of a Kerr–Newman black hole to the linear order of particle's spin are provided.
Our main focus is on the motion of a particle oscillating between the two turning points, which are the two outermost roots of the radial potential. This motion is restricted to the equatorial plane. 
There is also an induced motion along the polar coordinate. 
The parameters of conserved quantities are specified.
In Sec. \ref{sec3}, the orbital solutions are obtained analytically and their trajectories are plotted.
Section \ref{sec4} is devoted to providing an order-of-magnitude estimate of the gravitational-wave emission due to EMRIs, based on a simplified version of kludge methods.

This work is not intended to construct a complete waveform model.
Instead, our goal is to provide analytical orbital solutions that can be used as input for future waveform calculations.
We plan to achieve this by including the effects of radiation reaction and self-force in a future study. 
Concluding remarks are drawn in Sec. \ref{sec5}.
In Appendix \ref{appendixA}, the radial potential and its roots of a spinless particle in the exterior of a Kerr–Newman black hole are summarized from \cite{wang-2022}.

\section{The dynamical equations for a spinning particle around the Kerr–Newman black hole}
\label{sec2}
The dynamical equations of motion for a spinning particle in curved spacetime are governed by the Mathisson–Papapetrou equations in the pole-dipole approximation, where the degrees of freedom involve a  point mass and  spin, with a coupling between the curvature and the spin given by \cite{corinaldesi-1951, mathisson-2010a, mathisson-2010b},
\begin{align}
&\frac{Dp^{\mu}}{d\sigma_m}=-\frac{1}{2} {R^{\mu}}_{\nu\kappa\lambda}\dot{x}^{\nu}S^{\kappa\lambda},\label{p_eq}\\
&\frac{DS^{\mu\nu}}{d\sigma_m}=p^{\mu}\dot{x}^{\nu}-p^{\nu}\dot{x}^{\mu}, \label{s_eq}
\end{align}
where $D/d\sigma_m$ denotes a covariant derivative with respect to the proper time along the worldline, and ${R^{\mu}}_{\nu\kappa\lambda}$ is the Riemann curvature of the spacetime. Moreover $p^{\mu}$ and $\dot{x}^{\nu} (u^{\mu})$ are the 4-momentum and 4-velocity of the particle, where the dot denotes the derivative with respect to the proper time.
Additionally, $S^{\mu\nu}$ \footnote{The Greek letters $\mu, \nu, ...$ in the superscript/subscript represent $ t, r, \theta, \phi$.} is the spin tensor, obeying skew-symmetric $S^{\mu\nu}=-S^{\nu\mu}$.
The Tulczyjew-Dixon spin supplementary condition, $S^{\mu\nu}p_{\nu}=0$ in \cite{Tul_1959,dixon-1964} leads to
\begin{align}
 p^{\mu}=m\dot{x}^{\mu}+\mathcal{O}\left(s^2\right) ,
\end{align}
where $m$ is the particle's mass.
In \cite{kyrian-2007} the spin tensor  can  be expressed as
\begin{align}
&S^{\mu\nu}=m\epsilon^{\mu\nu\alpha\beta}\dot{x}_{\alpha}s_{\beta}\label{big S},
\end{align}
or equivalently
\begin{align}
&s^{\mu}=-\frac{1}{2} m {\epsilon^{\mu\nu}}_{\alpha\beta} \, \dot{x}_{\nu} S^{\alpha \beta} \label{small S},
\end{align}
where $s^{\mu}$ is the specific spin vector and $\epsilon^{\mu\nu\kappa\lambda}$ is the Levi-Civita pseudo-tensor.
See recent articles for details \cite{drummond-2022A, drummond-2022B}.
To the linear order of spin \cite{witzany-2024}, Eqs. (\ref{p_eq}) and (\ref{s_eq}) can be approximated by
\begin{align}
&\frac{D^2 x^{\mu}}{d \sigma_m^2}=-\frac{1}{2} {R^{\mu}}_{\nu\gamma\delta} {\epsilon^{\gamma\delta}}_{\kappa \lambda} \, \dot{x}^{\nu} \dot{x}^{\kappa} s^{\lambda}\, ,\label{p_eq_s}\,\\
&\frac{D s^{\lambda}}{d\sigma_m}=0\, .\label{s_eq_s}
\end{align}
Thus, the spin vector $s^\lambda$ is parallel-transported along geodesics.

The line element of the Kerr–Newman black hole exterior takes the form:
\begin{align}
ds^2=-\frac{\Delta}{\Sigma} \left(dt-a\sin^2\theta d\phi \right)^2+\frac{\sin^2\theta}{\Sigma} \left[(r^2+a^2)d\phi-adt \right]^2 +\frac{\Sigma}{\Delta}dr^2+\Sigma d\theta^2 ,
\label{Line element}
\end{align}
where
\begin{align}
&\Delta=r^2-2Mr+a^2+Q^2,\label{Delta}\\
&\Sigma=r^2+a^2\cos^2{\theta}\label{Sigma}
\end{align}
with the black hole's mass $M$, charge $Q$, angular momentum $J$, and angular momentum per unit mass $a=J/M$. 
The inner and outer horizons are given by
\begin{align}
    r_\pm = M \pm \sqrt{M^2 -(a^2 +Q^2)} .
\end{align}
For the case of $a=0$, or $a=0$, $Q=0$, the metric reduces to that of the Reissner–Nordstr\"om or Schwarzschild black hole.
The associated Killing vectors are generators of  translation in time and rotation of the azimuthal angle with respect to the black hole's spin given by
\begin{align}
&\xi^{\mu}_{t}=\delta^{\mu}_{t}\quad,\quad \xi^{\mu}_{\phi}=\delta^{\mu}_{\phi},
\end{align}
where the constants of motion, namely the energy $E_m$ \footnote{In \cite{ciou-2025}, $C_{(t)} \equiv E_m$, but we define $C_{(t)} \equiv -E_m$ here.} and azimuthal angular momentum $L_{m}$, can be obtained from the expression in \cite{dixon-1964}, which is
\begin{equation}
C_{(\xi)}\equiv p_{\mu}\xi^{\mu}-\frac{1}{2}(\nabla_{\sigma}\xi_{\rho})S^{\rho\sigma}. \label{const}
\end{equation}

To obtain them, we explicitly write down each component of the spin tensor \footnote{In \cite{ciou-2025}, the component $S^{r\phi}$ misses a minus sign.} from (\ref{big S}),
\begin{align}
\begin{cases}
    \ S^{tr} = -\frac{m \sin\theta}{\Sigma} \left\{ a[\Delta-(r^2+a^2)](\dot{\theta}s^{t}-\dot{t}s^{\theta})-[\Delta a^2\sin^2{\theta}-(r^2+a^2)^2](\dot{\theta}s^{\phi}-\dot{\phi}s^{\theta})\right\}
    =-S^{rt},\\
    \ S^{t\theta} = \frac{m \sin{\theta}}{\Delta \Sigma} \left\{ a[\Delta-(r^2+a^2)](\dot{r}s^{t}-\dot{t}s^{r})-[\Delta a^2\sin^2{\theta}-(r^2+a^2)^2](\dot{r}s^{\phi}-\dot{\phi}s^{r})\right\}
    =-S^{\theta t},\\
    \ S^{t\phi} = -\frac{m \Sigma}{\Delta \sin{\theta}}(\dot{r}s^{\theta}-\dot{\theta}s^{r})
    =-S^{\phi t},\\
    \ S^{r \theta} = \frac{m \Delta \sin{\theta}}{\Sigma}(\dot{t}s^{\phi}-\dot{\phi}s^{t})
    =-S^{\theta r},\\
    \ S^{r \phi} = -\frac{m }{\Sigma \sin{\theta}} \left\{ (\Delta-a^2\sin^2{\theta})(\dot{t}s^{\theta}-\dot{\theta}s^{t})-a\sin^2{\theta}[\Delta-(r^2+a^2)](\dot{\phi}s^{\theta}-\dot{\theta}s^{\phi})\right\}
    =-S^{\phi r},\\
    \ S^{\theta \phi} = \frac{m}{\Delta\Sigma \sin{\theta}} \left\{ (\Delta-a^2\sin^2{\theta})(\dot{t}s^{r}-\dot{r}s^{t})-a\sin^2{\theta}[\Delta-(r^2+a^2)](\dot{\phi}s^{r}-\dot{r}s^{\phi}) \right\}
    =-S^{\phi \theta},
\end{cases} \label{Smunu}
\end{align}
where $ S^{tt},S^{rr},S^{\theta\theta},S^{\phi \phi}$ are all zero.
The expressions for the constants of motion
\begin{align}
\gamma_m=\frac{E_m}{m}, \quad \lambda_m=\frac{L_m}{m}
\end{align}
can be derived from (\ref{const}) and (\ref{Smunu}) as
\begin{align}
&\gamma_m=\frac{1}{\Sigma}(\Delta-a^2\sin^2{\theta})\dot t -\frac{a\sin^2{\theta}}{\Sigma}[\Delta-(r^2+a^2)]\dot {\phi}\notag\\
&\quad\quad-\frac{\sin{\theta}}{\Sigma^2} \left[r(M r-Q^2)-Ma^2\cos^2{\theta} \right] \left[ a(\dot{\theta}s^{t}-\dot{t}s^{\theta})-(r^2+a^2)(\dot{\theta}s^{\phi}-\dot{\phi}s^{\theta}) \right] \notag\\
&\quad\quad-\frac{a\cos{\theta}}{\Sigma^2} (2M r-Q^2) \left[ \dot{r}s^{t}-\dot{t}s^{r}-a\sin^2{\theta}(\dot{r}s^{\phi}-\dot{\phi}s^{r}) \right], \label{energy}\\
&\lambda_{m}=\frac{a\sin^2{\theta}}{\Sigma} \left[ \Delta-(r^2+a^2) \right] \dot t -\frac{\sin^2{\theta}}{\Sigma} \left[ \Delta a^2\sin^2{\theta}-(r^2+a^2)^2 \right] \dot{\phi}\notag\\
&\quad\quad-\frac{\sin{\theta}}{\Sigma^2} \left[r(r^2+a^2)(2M r-Q^2)-M\Sigma a^2\sin^2{\theta}-r\Sigma^2 \right](\dot{\theta}s^{t}-\dot{t}s^{\theta})\notag\\
&\quad\quad-\frac{a\sin^3{\theta}}{\Sigma^2} \left( M\Sigma(r^2+a^2)-r(2M r-Q^2)[\Sigma+(r^2+a^2)] \right)(\dot{\theta}s^{\phi}-\dot{\phi}s^{\theta})\notag\\
&\quad\quad-\frac{\cos{\theta}}{\Sigma^2} \left( \Sigma^2+\Sigma a^2\sin^2{\theta}+a^4\sin^4{\theta}-\Delta a^2\sin^2 \right) (\dot{r}s^{t}-\dot{t}s^{r})\notag\\
&\quad\quad-\frac{a^3\sin^4{\theta}\cos{\theta}}{\Sigma^2} \left[ \Delta-(r^2+a^2) \right] (\dot{r}s^{\phi}-\dot{\phi}s^{r}), \label{orbit angular momentum}
\end{align}
where the first term of each of the above expressions is independent of spin, while the other terms provide spin corrections. 
In the limit of $a\rightarrow 0$, or $a\rightarrow 0$, $Q\rightarrow 0$, these expressions become those of the Reissner–Nordstr\"om \cite{ciou-2025} or Schwarzschild \cite{witzany-2024} black hole.
Now, we introduce an antisymmetric Killing-Yano tensor \cite{tanaka-1996},
\begin{align}
F_{\mu\nu} = a \cos{\theta} \left( \bar{e}_{\mu}^{(1)} \bar{e}_{\nu}^{(0)} - \bar{e}_{\mu}^{(0)} \bar{e}_{\nu}^{(1)} \right)+r \left( \bar{e}_{\mu}^{(2)} \bar{e}_{\nu}^{(3)} - \bar{e}_{\mu}^{(3)} \bar{e}_{\nu}^{(2)} \right), \label{killing yano tensor}
\end{align}
where
\begin{align}
\bar{e}_{\mu}^{(0)} &= \left(\sqrt{\frac{\Delta}{\Sigma}}, 0, 0, -a\sin^2{\theta}\sqrt{\frac{\Delta}{\Sigma}}\right), \\
\bar{e}_{\mu}^{(1)} &= \left(0, \sqrt{\frac{\Sigma}{\Delta}}, 0, 0\right), \\
\bar{e}_{\mu}^{(2)} &= \left(0, 0, \sqrt{\Sigma}, 0\right), \\
\bar{e}_{\mu}^{(3)} &= \left(-\frac{a\sin{\theta}}{\sqrt{\Sigma}}, 0, 0, \frac{(r^2+a^2)\sin{\theta}}{\sqrt{\Sigma}}\right) ,
\end{align}
obeying
\begin{equation}\label{F_sym}
\nabla_\gamma F_{\alpha\beta}+\nabla_\beta F_{\alpha \gamma}=0.
\end{equation}
We can define a vector from (\ref{killing yano tensor}) to be
\begin{align}
l^\nu &= F^{\mu\nu} \dot{x}_\mu,\label{l_4_vec}
\end{align}
where
\begin{equation} \label{l_0}
\begin{split}
&l^t=a \left[r\sin\theta
 \dot{\theta}-\frac{\cos\theta}{\Delta}(r^2+a^2)\dot{r} \right]\,, \\
&l^r=\frac{\Delta}{\Sigma}a\cos\theta(a\sin^2\theta \dot{\phi}-\dot{t})\,, \\
&l^\theta=\frac{r}{\Sigma}\sin\theta[a\dot{t}-(r^2+a^2)\dot{\phi})]\,, \\
&l^\phi =\frac{r}{\sin\theta}\dot{\theta}-\frac{a^2}{\Delta}\cos\theta\dot{r}\,.
\end{split}
\end{equation}
This is called the orbital angular momentum 4-vector \cite{drummond-2022A}.
Since $l^\nu$ in (\ref{l_4_vec}) is parallel-transported along geodesics in general Kerr–Newman spacetimes resulting from (\ref{F_sym}), for a spinning particle following the non-geodesics, ${D l^{\mu}}/{d\sigma_m} \propto \mathcal{O}(s)$.
Thus, with (\ref{s_eq_s}), the component of the spin vector $s_{\lVert}$ is an approximate constant of motion
\begin{align}
&s_{\lVert}\equiv\frac{l^{\mu}s_{\mu}}{\sqrt{K}}, \label{s_parallel_def}\\
&\frac{ds_{\lVert}}{d\sigma_m}=0+\mathcal{O}(s^2) \label{s_parallel}
\end{align}
with
\begin{equation}
K \equiv l^{\mu}l_{\mu}.\label{K}
\end{equation}
Given the constants of motion $\gamma_{m}$ and $\lambda_m$ together with $\dot x^{\mu} \dot x_{\mu}=-1$, the equations of motion for the orbital variables $r(\sigma_m)$, $t(\sigma_m)$, and $\phi(\sigma_m)$ to the first order in $s_\parallel$ with $s^\theta = -s_{\parallel}/r $ given by (\ref{l_0}) and (\ref{s_parallel_def}) at $\theta=\pi/2$ are obtained as
\begin{align}
&r^2 \frac{d r}{d \sigma_m}=\pm_{r} \sqrt{{{ R}^s_m}(r)} \, ,\label{r_eq}\\
&r^2 \frac{d t}{d\sigma_m}=\frac{r^2+a^2}{\Delta} \left[ (r^2+a^2)\gamma_m -a\lambda_m \right] -a(a\gamma_m-\lambda_m)\notag\\
&\quad\quad\quad\quad+\frac{s_{\parallel}}{\Delta} \left[ a \left( 3M r-2Q^2+a^2\frac{M r-Q^2}{r^2} \right) \gamma_m-(r^2+a^2)\frac{M r-Q^2}{ r^2}\lambda_m \right], \label{t eq}\\
&r^2 \frac{d\phi}{d \sigma_m}=\frac{a}{\Delta} \left[ (r^2+a^2)\gamma_m-a\lambda_m \right]-(a\gamma_m-\lambda_m)\notag\\
&\quad\quad\quad\quad+\frac{s_{\parallel}}{\Delta} \left[ \left( 2M r-Q^2-r^2+a^2 \frac{M r-Q^2}{r^2} \right) \gamma_m-a\frac{M r-Q^2}{ r^2}\lambda_m \right] , \label{phi eq}
\end{align}
where the radial potential $R_m^s (r)$ is given by
\begin{equation}
\begin{split}
{R^s_m}(r) \equiv & \, \left[(r^2+a^2)\gamma_{m}-a\lambda_{m} \right]^2-\Delta \left[(a\gamma_{m}-\lambda_{m})^2+r^2 \right]\\
&+2 s_\parallel \left[\gamma_m \lambda_m r^2+\gamma_m (a \gamma_m-\lambda_m)(3M r-2Q^2)+ a (a \gamma_m-\lambda_m)^2 \frac{M r-Q^2}{r^2} \right]. \label{R}
\end{split}
\end{equation}
Here, to obtain the above  equations of motion with the correction of $s_{\parallel}$ has involved the assumption that $s_\parallel > \sqrt{s^2 -s_\parallel^2} $.
Moreover, the dynamics of the misaligned spin vector with $s_\parallel \neq s$ induces the motion of the particle out of the equatorial plane denoted by
\begin{equation}
\theta(\tau_m) = \frac{\pi}{2} + \delta\theta(\tau_m) ,
\end{equation}
which satisfies the Mathisson–Papapetrou equations (\ref{p_eq}). In Mino time, defined by $\frac{dx^{\mu}}{d\tau_m}=r^2\frac{dx^{\mu}}{d\sigma_m}$, the equation is obtained to be
\begin{footnotesize}
\begin{equation}
\begin{split}
\frac{d^2\delta\theta}{d\tau_m^2} &
-\frac{ a^4 \gamma_m^2 \left(Q^2-2 M r\right)
-2 a^3 \gamma_m \lambda_m \left(Q^2-2 M r\right)
-\lambda_m^2 r^2 \left(\Delta-a^2\right)
+a^2 \left[\left(Q^2-2 M r\right) \left(\lambda_m^2+\gamma_m^2 r^2\right) +({U^r})^2 \right]}{r^2 \Delta} \delta\theta \\
&= \frac{(2 Q^2 -3M r) \sqrt{K} \left[ a^3 s^\phi {U^r}+a^2 \left(\gamma_m r^2 s^r-s^t {U^r}\right)+a s^\phi {U^r} \left(\Delta-a^2\right)-a \lambda_m r^2 s^r-s^t {U^r} \left(\Delta-a^2\right)+\gamma_m r^4 s^r \right]}{r^4 \Delta}\,  \label{nongeodesic_theta}
\end{split}
\end{equation}
\end{footnotesize}
with $U^r=r^2 \dot r$, where $s^t$, $s^r$, and $s^\phi$ depend on $\sqrt{s^2 -s_\parallel^2} $ to be discussed later.
This equation reduces to Eq. (5.15) of \cite{drummond-2022A} for the Kerr case in the limit of $Q\rightarrow 0$. 

Because the spin degrees of freedom obey (\ref{s_eq_s}), to express them in the linear spin regime, we introduce the Marck tetrad $e_{(\alpha)}^\mu$ for $\alpha =0,1,2,3$ in \cite{marck-1983, van-de-meent-2020}, which are orthogonal each other and also parallel transported along geodesics.
We start form the first leg $e_{(0)}^\mu$ to be the geodesic's 4-velocity given by (\ref{r_eq}), (\ref{t eq}), and (\ref{phi eq}) with $s_{\lVert}=0$, that is,
\begin{equation}
\begin{split}
{e}_{(0)}^t &= \frac{r^2+a^2}{r^2 \Delta} \left[ (r^2+a^2)\gamma_m -a\lambda_m \right] -\frac{a(a\gamma_m-\lambda_m)}{r^2}, \\
{e}_{(0)}^r &= \pm_r \frac{1}{r^2}\sqrt{\mathcal{R}_m}, \\
{e}_{(0)}^\theta &= 0 , \\
{e}_{(0)}^\phi &= \frac{a}{r^2 \Delta} \left[ (r^2+a^2)\gamma_m-a\lambda_m \right] -\frac{(a\gamma_m-\lambda_m)}{r^2}\, ,
\end{split}
\end{equation}
where $\mathcal{R}_m$ is the radial potential for a spinless particle defined in Appendix \ref{appendixA}.
And the second leg $e_{(3)}^\mu$ can be the normalized $l^\mu$ in (\ref{l_4_vec}), namely $e_{(3)}^\mu = l^\mu/ \sqrt{K}$ in the case of $\theta = \pi/2$ and $\dot \theta =0 $, where $K$ in (\ref{K}) becomes $K=(a \gamma_m -\lambda_m)^2$. 
Now, the second leg $e_{(3)}^\mu$ is
\begin{align}
e^{\mu}_{(3)} &= \left(0,0, -\frac{1}{r}, 0 \right).
\label{tetrad}
\end{align}
Then, two other tetrads, which are orthogonal to the plane of $e_{(0)}^\mu$, $e_{(3)}^\mu$ and orthogonal to each other can be constructed from $\tilde{e}_{\mu(1)}$ and $\tilde{e}_{\mu(2)}$,
\begin{align}
\begin{cases}
\tilde{e}_{t(1)} = -\frac{r}{\sqrt{K + r^2}}\dot{r} \\
\tilde{e}_{r(1)} = \frac{r}{\Delta\sqrt{K + r^2}} \left[ (r^2+a^2)\gamma_m-a\lambda_m \right] \\
\tilde{e}_{\theta(1)} = 0 \\
\tilde{e}_{\phi(1)} = \frac{a r}{\sqrt{K + r^2}}\dot r \\
\end{cases},
\begin{cases}
\tilde{e}_{t(2)} = \sqrt{\frac{K + r^2}{K}} \left\{ \gamma_m -\frac{1}{r^2} \left( 1-\frac{K}{K+r^2} \right) \left[ (r^2+a^2)\gamma_m-a\lambda_m \right] \right\} \\
\tilde{e}_{r(2)} = -\frac{r^2}{\Delta}\sqrt{\frac{K}{K + r^2}}\dot{r} \\
\tilde{e}_{\theta(2)} = 0 \\
\tilde{e}_{\phi(2)} =\frac{1}{r^2} \sqrt{\frac{K + r^2}{K}} \left( \frac{K}{K+r^2}-1\right) (r^2+a^2) (\lambda_m-a\gamma_m) -\sqrt{\frac{K}{K + r^2}} \lambda_m
\end{cases} \label{tetrad1}
\end{align}
in the orbital rotation plane.
Then $e_{(1)}^\mu$ and $e_{(2)}^\mu$ are obtained by linear combinations of $\tilde{e}_{(1)}^\mu$ and $\tilde{e}_{(2)}^\mu$ with an introduced precession angle $\psi$ to be
\begin{align}
e_{(1)}^\mu&=\tilde{e}^{\mu}_{(1)}\cos{\psi}+\tilde{e}^{\mu}_{(2)}\sin{\psi},\\
e_{(2)}^\mu&=-\tilde{e}^{\mu}_{(1)}\sin{\psi}+\tilde{e}^{\mu}_{(2)}\cos{\psi}.
\end{align}
The requirement of $e_{(1)}^\mu$ and $e_{(2)}^\mu$ being parallel-transported along geodesic leads to the equation for the precession angle $\psi(\sigma_m)$,
\begin{align}
\frac{d \psi}{d \tau_m}={\sqrt{K}}\left[\frac{(r^2+a^2)\gamma_m-a\lambda_m}{K+r^2}+\frac{a}{K}(\lambda_m-a\gamma_m)\right]
\end{align}
which is consistent with (55) in \cite{van-de-meent-2020} when $\theta =\pi/2$.
We rewrite the spin vector in the basis of the tetrad, namely $s^\mu=s^{(0)}e_{(0)}^{\mu} +s^{(1)}e_{(1)}^{\mu} +s^{(2)}e_{(2)}^{\mu} +s^{(3)}e_{(3)}^{\mu}$.
Following \cite{ciou-2025}, the spin supplement condition requires that $u_\alpha s^\alpha =0$, leading to $s^{(0)} =0$ due to $e_{(0)}^\mu =u^\mu$.
According to $s^\theta = -s_{\parallel}/r $ and $e_{(3)}^\mu$ in (\ref{tetrad}), $s^{(3)} = s_\parallel$ \footnote{In \cite{ciou-2025}, it should be $s^{(3)} = s_\parallel$. Therefore, the spin vector is given by $s^\mu = s_\perp {e}^{\mu}_{(2)} +s_\parallel {e}^{\mu}_{(3)}$.}.
For $s^2 = s_{\parallel}^2 +s_{\perp}^2$, we can write  $s^{(1)}=s_\perp \cos{w}$, $s^{(2)} =s_\perp \sin{w}$.
The spin vector can be written as
\begin{footnotesize}
\begin{equation} \label{spinvector}
\begin{split}
s^t& = s_\perp
\frac{\sqrt{K} r  U^r \left(a^2+r^2\right)\cos{w} + \left\{a^2 \gamma_m \left[ K \left(Q^2-r (2 M+r)\right)+\Delta  r^2\right] -a \lambda_m \left[K \left(Q^2-2 M r\right)+\Delta  r^2\right] -\gamma_m K r^4\right\} \sin{w} }{r^2 \Delta \sqrt{K \left(K+r^2\right)}} \cos{\psi} \\
+& s_\perp
\frac{-\sqrt{K} r  U^r \left(a^2+r^2\right)\sin{w} + \left\{a^2 \gamma_m \left[K \left(Q^2-r (2 M+r)\right)+\Delta  r^2\right] -a \lambda_m \left[K \left(Q^2-2 M r\right)+\Delta  r^2\right] -\gamma_m K r^4\right\}\cos{w} }{r^2 \Delta \sqrt{K \left(K+r^2\right)}} \sin{\psi},\\
s^r& = s_\perp \frac{ r \left[\gamma_m \left(a^2+r^2\right)-a \lambda_m\right] \cos{(\psi+w)} -\sqrt{K} U^r \sin{(\psi+w)} }{r^2 \sqrt{K+r^2}},\\
s^{\theta}& = -\frac{s_\parallel}{r}, \\
s^{\phi}& = s_\perp
\frac{a \sqrt{K} r  U^r \cos{w} + \left\{ a \gamma_m  \left[r^2 \Delta+K \left(Q^2-2 M r\right)\right] -\lambda_m  \left[r^2 \Delta +K \left(\Delta-a^2\right)\right] \right\} \sin{w} }{r^2 \Delta \sqrt{K \left(K+r^2\right)}} \cos{\psi} \\
&+ s_\perp
\frac{ -a \sqrt{K} r  U^r \sin{w} + \left\{ a \gamma_m  \left[r^2 \Delta +K \left(Q^2-2 M r\right)\right] -\lambda_m  \left[r^2 \Delta +K \left(\Delta-a^2\right)\right] \right\} \cos{w} }{r^2 \Delta \sqrt{K \left(K+r^2\right)}} \sin{\psi} .
\end{split}
\end{equation}
\end{footnotesize}
Then, substituting the above spin vector (\ref{spinvector}) and the radial velocity (\ref{r_eq}) into (\ref{nongeodesic_theta}), the equation of motion for the induced angle $\delta\theta$ becomes
\begin{equation} \label{deltatheta_eq}
\frac{d^2\delta\theta}{d\tau_m^2} + \left[ \omega_{\delta \theta, 0}^B \right]^2 \delta\theta
= \sqrt{s^2 -s_\parallel^2}\; \frac{(2 Q^2 -3M r) (\lambda_m -a \gamma_m) \sqrt{r^2 +K}}{r^3} \cos{(\psi +w)}
\end{equation}
with the polar frequency
\begin{equation}
\omega_{ {\delta \theta, 0}}^B = \sqrt{\lambda_m^2 +a^2 \left(1-\gamma_m^2\right) } \,,
\end{equation}
which reduces to Eq. (3.77) in \cite{piovano-2025A} in the case of $Q=0$. 
The value of the parameter $w$ depends on the  choice of the initial condition of $\psi$. 
In addition, the initial conditions of $\delta\theta$ can be chosen so that the homogeneous solution vanishes \cite{witzany-2024,ciou-2025}.
The inhomogeneous solution was proposed in \cite{piovano-2025A} for the Kerr black hole  as
\begin{equation} \label{deltatheta2}
\delta\theta(\tau_m)
= -\sqrt{s^2 -s_\parallel^2}\; \frac{\sqrt{r^2 +K}}{r \,(\lambda_m -a \gamma_m)} \cos{(\psi +w)} \, .
\end{equation}
In the limit of $a\rightarrow 0$ with $w=\pi/2$, the solution reduces not only to that of the Schwarzschild black hole \cite{witzany-2024} but also to the Reissner–Nordstr\"om black hole with an additional black hole's charge \cite{ciou-2025}.
Thus, a novel result is found that the solution $\delta\theta$ in (\ref{deltatheta2}) also holds true for the Kerr–Newman black hole. 
Using (\ref{psim}), (\ref{Ipsi}), (\ref{I_psi}), (\ref{I_psi0}), and the fact that $\frac{d r}{d \tau_m}=\pm_{r} \sqrt{{\mathcal{R}}_m(r)}$ defined in Appendix \ref{appendixA}, plugging (\ref{deltatheta2}) into (\ref{deltatheta_eq}) and using the \texttt{Mathematica} codes in \cite{chen_2026_21881833} can directly verifies  that the inhomogeneous solution satisfies the equation of motion.
Note that although this solution apparently does not explicitly depend on the charge $Q$, on the way to justify the solution, the dependence of $Q$ is introduced through $\frac{dr}{d \tau_m}$, where the $Q$-dependent radial potential is involved. 

Before solving the equations of motion to show the analytical solutions, we next study the roots of the radial potential.
In general, there are six roots of the radial potential,
\begin{align}
&R^s_m(r) = \frac{1}{r^2}(\gamma_{m}^2 - 1) (r - r_{m1})(r - r_{m2})(r - r_{m3})(r - r_{m4})(r - r_{m5})(r - r_{m6}). \label{R four root}
\end{align}
Four of them are modified from the corresponding roots for a spinless particle.
To the first order in $s_\parallel$, they can be written as
\begin{align}\label{roots}
r_{mj} = r_{mj}^{(0)} +s_\parallel r_{mj}^{(1)}
\end{align}
with $j=1,2,3,4$, where $r_{mj}^{(0)}$ are the four roots of (\ref{R}) when $s_\parallel = 0$ in Appendix \ref{appendixA}.
The small modification from the spin $s_\parallel$ is given by
\begin{align}
r_{mj}^{(1)} =\frac{\gamma_{m} \lambda_{m} r_{mj}^{(0)4}+\gamma_{m} (a \gamma_{m} -\lambda_{m} ) \left(3M r_{mj}^{(0)}-2Q^2 \right) r_{mj}^{(0)2} +a (a \gamma_{m}-\lambda_{m})^2 \left(M r_{mj}^{(0)}-Q^2\right)}{2 \left(\gamma_{m}^2-1\right) r_{mj}^{(0)5}+3 M r_{mj}^{(0)4}+[a^2 \left(\gamma_{m}^2-1\right)-\lambda_{m}^2-Q^2]r_{mj}^{(0)3}+M r_{mj}^{(0)2} (\lambda_{m}-a \gamma_{m})^2}
\label{root1}
\end{align}
which is consistent with \cite{ciou-2025} for $M s_\parallel/Q^2 \to 0$.
Additionally, the fifth and sixth roots with $j=5,6$, which appear in pairs, can be shown in the following  limits.
In the limit of $M s_\parallel/Q^2 \to 0$, the roots are
\begin{align}
&r_{m5} \to -i \sqrt{2 a} \sqrt{s_\parallel} -\frac{a M s_\parallel}{Q^2} + \frac{9 i a^{3/2} M^2 s_\parallel^{3/2}}{2 \sqrt{2} Q^4} \,,\\
&r_{m6} \to i \sqrt{2a} \sqrt{s_\parallel} -\frac{a M s_\parallel}{Q^2} - \frac{9 i a^{3/2} M^2 s_\parallel^{3/2}}{2 \sqrt{2} Q^4} \,.
\end{align}
To simplify orbital solutions, we take into account the first term, namely
\begin{equation} \label{r5r6}
r_{mj} = r_{mj}^{(1)} \sqrt{s_\parallel}\, ,
\end{equation}
 where
\begin{align}
&r_{m5}^{(1)} \to -i \sqrt{2 a}  \,,\quad r_{m6}^{(1)} \to i \sqrt{2a}  \,
\end{align}
in the following integrals.
For $s_\parallel > 0$, they are a complex-conjugated pair, i.e. $r_{m5} = r^*_{m6}$. However, for $s_\parallel < 0$, they become two real-valued roots, i.e. $r_{m5} = - r_{m6}$. 
To produce the plots below, we consider the limit of $M s_\parallel/Q^2 \to 0$ and $s_\parallel>0$. 
For comparison with the result in the Kerr case \cite{piovano-2025B}, we also consider another limit $Q^2/(M s_\parallel) \to 0$, where the roots become
\begin{align}
&r_{m5} \to -i \sqrt{a} \sqrt{s_\parallel} -\frac{Q^2}{4 M} -\frac{7 i Q^4}{32 \sqrt{a} M^2 \sqrt{s_\parallel}} \,,\\
&r_{m6} \to i \sqrt{a} \sqrt{s_\parallel} -\frac{Q^2}{4 M} + \frac{7 i Q^4}{32 \sqrt{a} M^2 \sqrt{s_\parallel}} \, .
\end{align}

\begin{figure}[htp]
\begin{center}
    \includegraphics[width=13cm]{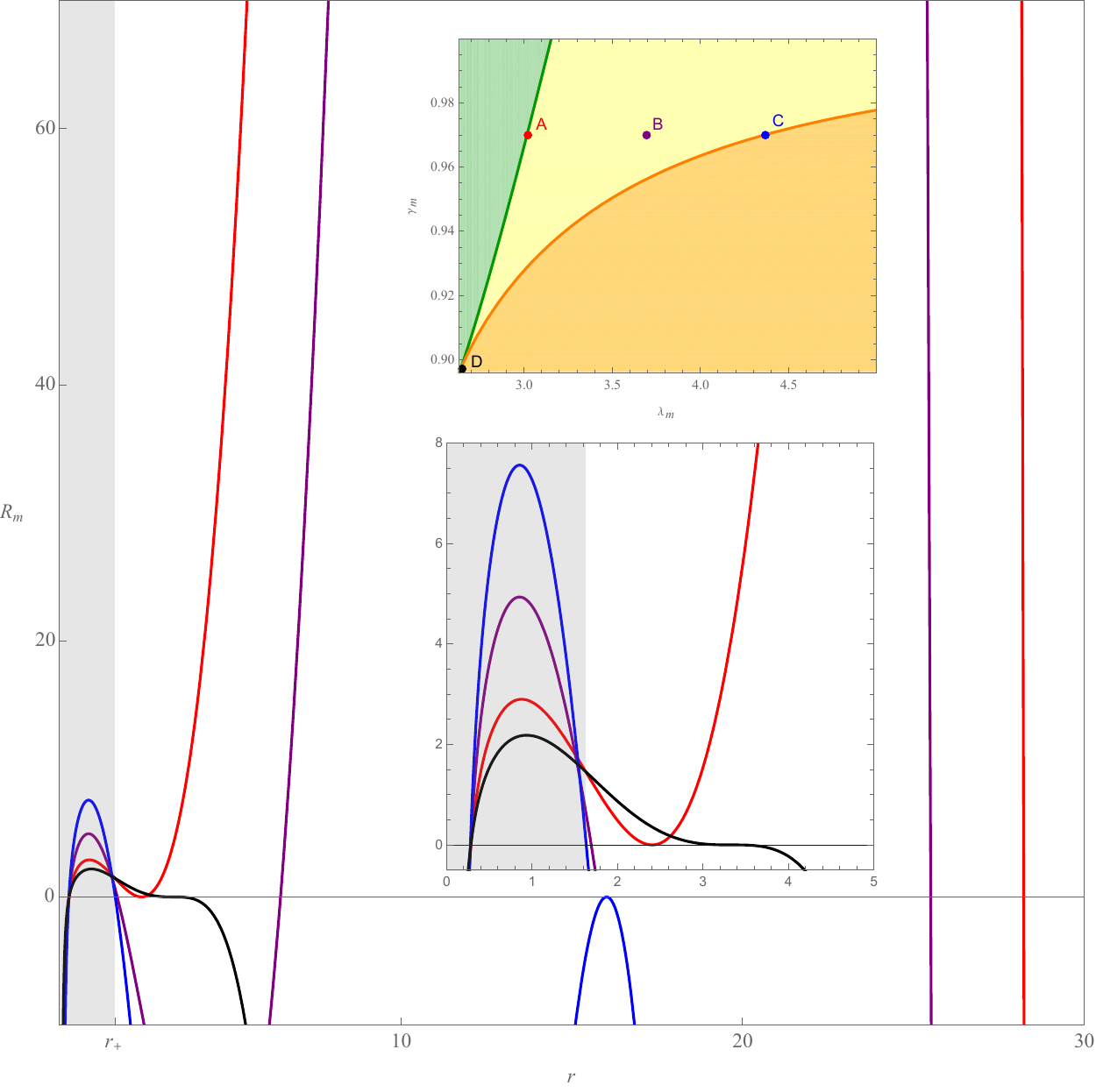}
    \caption{
    The plot shows the radial potential $R_m^s(r)$ and the parameter space $(\lambda_m, \gamma_m)$ for black hole spin $a/M = 0.5$, charge $Q/M = 0.6$, and particle spin $s_\parallel/m = 0.1$ in the case of the bound motion for $\gamma_m<1$.
    In the right inset, the green (orange) line represents the parameters for the unstable (stable) double root of $r_{m2} =r_{m3}$ ($r_{m3} =r_{m4}$) with the representative point A(C).
    Two double roots merge at point D of a triple root.
    The parameters at point B cause a particle to oscillate between the turning points $r_{m3}$ and $r_{m4}$.
    The lower inset plot shows the details of the radial potential in the main figure with the parameters at points A-D. }
\label{KN radial potential and parameter space}
\end{center}
\end{figure}

\begin{figure}[htp]
\begin{center}
    \includegraphics[width=13cm]{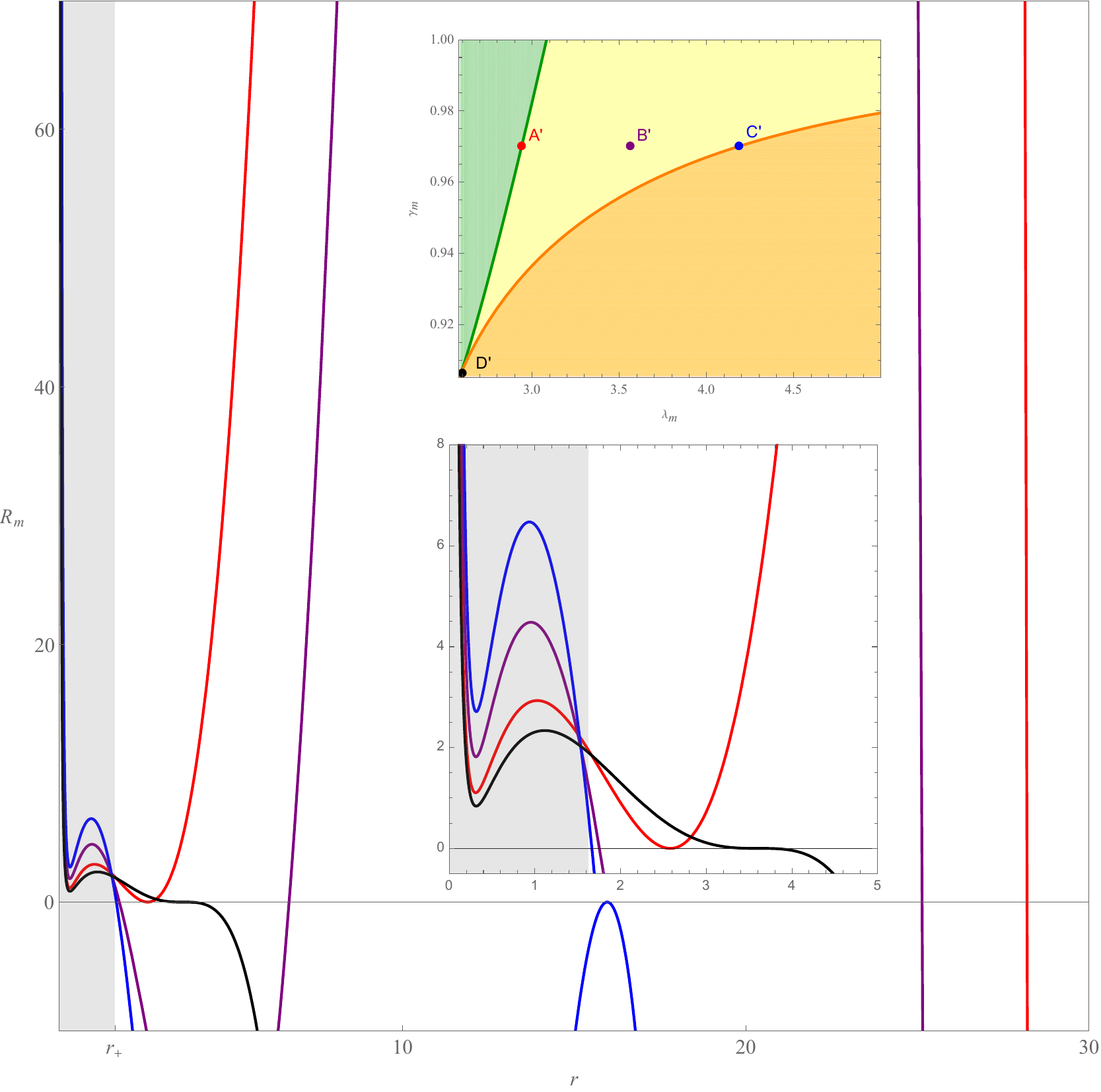}
    \caption{
    This plot shows the radial potential $R_m^s(r)$ and the parameter space $(\lambda_m, \gamma_m)$ for the same black hole parameters as in Fig. \ref{KN radial potential and parameter space} with an anti-parallel spin $s_\parallel/m = -0.1$ to the orbital angular momentum.
    In the right inset, the green (orange) line represents the parameters for the unstable (stable) double root of $r_{m2} =r_{m3}$ ($r_{m3} =r_{m4}$) with the representative point A$'$ (C$'$).
    Two double roots merge at point D$'$ of a triple root.
    The parameters at the point B$'$ cause a particle to oscillate between the turning points $r_{m3}$ and $r_{m4}$.
    The radial potential is shown for the parameters at points A$'$-D$'$. }
\label{KN radial potential and parameter space 2}
\end{center}
\end{figure}

In Fig. \ref{KN radial potential and parameter space} with $s_\parallel >0$, the orange line shows the parameters for the stable double root of $r_{m3}=r_{m4}$ with the representative parameters at point C.
The green line shows the parameters for the unstable double root of $r_{m2}=r_{m3}$ with the representative parameters at point A.
Two lines merge at point D of a triple root ($r_{m2} =r_{m3} =r_{m4}$) with the parameters, leading to the innermost circular motion (ISCO).
They will be studied elsewhere.
The parameters of point B are chosen, allowing us to study the equatorial motion of a particle oscillating between two turning points $r_{m3}$ and $r_{m4}$.
The similar situation can be seen in Fig. \ref{KN radial potential and parameter space 2} with $s_\parallel <0$. 
The two extra real roots $r_{m5}$ and $r_{m6}$ affect the behavior of the radial potential inside the horizon.

\section{Equatorial motion and analytical solution }
\label{sec3}
For the source point $x^{\mu}_i$ and the observer point $x^{\mu}$, the integral forms of the equations of motion in Mino time $\tau_m$ can be written as
\begin{align}
\tau_m(r) -\tau_{mi}&=I_r, \label{taum}
\end{align}
and
\begin{align}
t(r) -t_{mi}&=I_t, \label{tm}\\
\phi(r) -\phi_{mi}&=I_\phi, \label{phim}\\
\psi(r) -\psi_{mi}&=I_\psi, \label{psim}
\end{align}
where
\begin{align}
I_r& \equiv \int_{r_i}^r\frac{dr'}{\sqrt{R_{m}^s(r')}} \,,\label{Ir}\\
I_t& \equiv \int_{r_i}^r\frac{dt}{d\tau_m}\frac{dr'}{\sqrt{R_{m}^s(r')}} \,, \label{It}\\
I_\phi& \equiv \int_{r_i}^r\frac{d\phi}{d\tau_m}\frac{dr'}{\sqrt{R_{m}^s(r')}} \,,\label{Iphi}\\
I_\psi& \equiv \int_{r_i}^r\frac{d\psi}{d\tau_m}\frac{dr'}{\sqrt{R_{m}^s(r')}} \,.\label{Ipsi}
\end{align}
According to \cite{witzany-2024,ciou-2025}, the solutions can be written in terms of a reference trajectory with corrections arising from the $s_\parallel$-dependent effects in the dynamical equations.
We now expand the integrand as
\begin{equation} \label{R_split}
\frac{1}{\sqrt{R^s_{m}(r)}}= \frac{1}{\sqrt{R_{m}(r)}} -s_{\parallel} \frac{r_{m5}^{(1)}r_{m6}^{(1)}}{2 {r'}^2 \sqrt{R_{m}(r)}} +\mathcal{O}(s_{\parallel}^2) \,,
\end{equation}
where
\begin{align}
R_{m}(r)= (\gamma_{m}^2 - 1) (r - r_{m1})(r - r_{m2})(r - r_{m3})(r - r_{m4}) .
\end{align}
The first term contributes to the motion involving the radial potential with the spin-corrected roots $r_{m1}$, $r_{m2}$, $r_{m3}$, and $r_{m4}$ given in (\ref{roots}).
The second term arises from the contributions of the roots of $r_{m5}^{(1)}$, $r_{m6}^{(1)}$ in (\ref{r5r6}).
The resulting solutions are valid up to linear order in spin, although they also retain higher order spin effects as in \cite{witzany-2024, ciou-2025}. 

As in previous studies \cite{wang-2022, ciou-2025}, the integrals can be decomposed into several functions: $I_0$, $I_1$, $I_2$, $I_{\pm}$, and $I_{i\pm}$.
Two additional functions, $I_{-1}$ and $I_{-2}$, are introduced due to the interaction between the particle's spin and the black hole's spin.
These functions are defined as
\begin{align}
I_n(r) &\equiv \sqrt{1-\gamma_m^2} \int_{r_i}^{r} \frac{(r')^n}{\pm_r \sqrt{R_m(r')}}dr' \quad\quad\quad\quad\quad {\rm for}\;{n=0,\pm 1, \pm 2}\, ,\\
I_\pm (r) &\equiv \sqrt{1-\gamma_m^2} \int_{r_i}^{r} \frac{dr'}{\pm_r\left(r'-r_{\pm} \right)\sqrt{R_m(r')}} \,, \\
I_{i\pm} (r) &\equiv \sqrt{1-\gamma_m^2} \int_{r_i}^{r} \frac{dr'}{\pm_r \left(r'\mp i \sqrt{K} \right) \sqrt{R_m(r')}} \,,
\end{align}
where the integral $I^B_{i-}$ is the complex conjugate of the integral $I^B_{i+}$.
The integrals will be computed later.\\

Now, we are ready to study the bound motion of a particle $(\gamma_m^2<1)$, which oscillates between two real-valued roots $r_{m3}<r<r_{m4}$ with the parameters of point B shown in Fig. \ref{KN radial potential and parameter space}.
The solution is denoted by the superscript "$B$".
Following (\ref{R_split}), we decompose (\ref{Ir}) as
\begin{align}
\tau_m(r) \equiv I_r^B(r) &= \int_{r_i}^r \left[ \frac{1}{\sqrt{R_{m}(r')}} -s_{\parallel} \frac{r_{m5}^{(1)}r_{m6}^{(1)}}{2 {r'}^2 \sqrt{R_{m}(r')}} +\mathcal{O}(s_{\parallel}^2) \right] dr' \notag\\
&= \frac{1}{\sqrt{1-\gamma_{m}^2}} I_0^B(r) -s_{\parallel} \frac{r_{m5}^{(1)}r_{m6}^{(1)}}{2 \sqrt{1-\gamma_{m}^2}} I_{-2}^B(r) +\mathcal{O}(s_{\parallel}^2) \,,  \label{I_r^B}
\end{align}
{where
\begin{align}
I_0^B(r) &= \frac{2 \nu_{r_{i}}}{\sqrt{(r_{m3}-r_{m1})(r_{m4}-r_{m2})}} F\left[\sin^{-1}\left(\sqrt{\frac{(r-r_{m3})(r_{m4}-r_{m2})}{(r-r_{m2})(r_{m4}-r_{m3})}}\right) \Bigg|k^B \right] -{\mathcal{I}^B_{0_i}} \label{IB0}
\end{align}
with
\begin{align}
k^B &=\frac{(r_{m4}-r_{m3})(r_{m2}-r_{m1})}{(r_{m3}-r_{m1})(r_{m4}-r_{m2})}\, .
\end{align}
The function $F(\varphi|k)$ is the incomplete elliptic integral of the first kind, whereas $I_{-2}^B$ will be shown below. }
Notice that $I_0^B$ takes the same form as the orbital solutions for a spinless particle around the Kerr–Newman black hole in \cite{wang-2022} but with the four $s_{\parallel}$-corrected roots in (\ref{roots}).
The corrections originate from the contributions of the roots of $r_{m5}^{(1)}$, $r_{m6}^{(1)}$ in (\ref{r5r6}) due to the spin effect, and  vanish when $s_{\parallel}=0$.

Using the properties of the Jacobian elliptic function \cite{van-de-meent-2020}, namely
${\rm sn}\left[ F(\sin^{-1}\varphi|k) |k \right] =\varphi$ and
${\rm am}\left[ F(\sin^{-1}\varphi|k) |k \right] =\sin^{-1}\varphi$, the function $\tau_m(r)$ in (\ref{I_r^B}) can be perturbatively inverted to obtain $r(\tau_m)$ in the following form
\begin{align}\label{r_o}
r(\tau_m)=\frac{r_{m3}(r_{m4}-r_{m2})-r_{m2}(r_{m4}-r_{m3}) \,{\rm sn}^2\left(X^{B}(\tau_m)\left|{k^{B}}\right)\right.}{(r_{m4}-r_{m2})-(r_{m4}-r_{m3}) \,{\rm sn}^2\left(X^{B}(\tau_m)\left|{k^{B}}\right)\right.} \,
\end{align}
with the perturbed phase  $X^{B} = X^{B(0)} +s_\parallel X^{B(1)}$, where
\begin{align}
X^{B(0)}(\tau_m) &= \frac{\sqrt{(1-\gamma_m^2)(r_{m4}-r_{m2})(r_{m3}-r_{m1})}}{2} \tau_m
+\nu_{r_i} F\Bigg[\sin^{-1}\left(\sqrt{\frac{(r_{m4}-r_{m2})(r_i-r_{m3})}{(r_{m4}-r_{m3})(r_i-r_{m2})}}\right)\left|{k^{B}}\Bigg]\right.\, , \label{X0_tau}\\
X^{B(1)}(\tau_m) &= \frac{\sqrt{(r_{m4}-r_{m2})(r_{m3}-r_{m1})} \, r_{m5}^{(1)} r_{m6}^{(1)}}{4} I_{-2}^B\left[r^{(0)}(\tau_m)\right] \, .\label{X1_tau}
\end{align}
The zeroth order term $ r^{(0)}(\tau_m)$ is given by (\ref{r_o}) with $X^{B(0)}(\tau_m)$ and has been shown in \cite{wang-2022, ciou-2025}.
Note that ${\rm sn}(\varphi|k)$ is the Jacobi elliptic sine function where ${\rm sn}^2(\varphi|k) \in [0,1]$ guarantees $r(\tau_m) \in [r_{m3}, r_{m4}]$.
The correction $s_\parallel X^{B(1)}$ provides a phase shift, which will be discussed in (\ref{periodR}). 

Then, given (\ref{r_o}), one can write the solutions of the orbital variables in terms of the coordinate $r$ as
\begin{align}
I_t^B(r) =&\, I_t^{B(0)}(r) +s_\parallel I_t^{B(1)}(r) ,\label{I_t}\\
I_\phi^B(r) =&\, I_\phi^{B(0)}(r) +s_\parallel I_\phi^{B(1)}(r) .\label{I_phi}
\end{align}
Again, the solution can be interpreted as the reference motion plus the correction.
The solution to the reference motion has the same form as the solution for a spinless particle around a Kerr–Newman black hole in \cite{wang-2022}, but it has spin-corrected roots in the radial potential.  This is not a geodesic motion because the solution involves the effect of spin $s_{\parallel}$ in the roots according to (\ref{p_eq_s}). The correction comes from the contributions due to the roots of $r_{m5}^{(1)}$ and $r_{m6}^{(1)}$ in (\ref{r5r6}).
Later, although the solution to the reference motion involves higher-order spin effects  as in \cite{witzany-2024, ciou-2025}, we will compare our analytical solutions with numerical integrations of the Mathisson–Papapetrou equations to confirm their accuracy in the linear spin order.
They are expressed as 
\begin{align}
I_t^{B(0)}(r) =&\, \frac{\gamma_m (4 M^2-Q^2)}{\sqrt{1-\gamma_{m}^2}} I_0^B + \frac{\gamma_m}{\sqrt{1-\gamma_m^2}} \left(I_2^B+2I_1^B M \right) \notag\\
&+\frac{1}{\sqrt{1-\gamma_m^2}(r_+-r_-)} \sum_{\xi=\pm} \xi (2 M r_\xi-Q^2) \left[ 2M\gamma_m r_\xi- ( a \lambda_m +\gamma_m Q^2 ) \right] I_\xi^B ,\label{I_t0}\\
I_t^{B(1)}(r) =&\, \frac{a \gamma_m-\lambda_m}{\sqrt{1-\gamma_m^2}} \left[MI_{-1}^B+(2 M^2-Q^2)I_{-2}^B \right]
- \frac{r_{m5}^{(1)} r_{m6}^{(1)}}{2\sqrt{1-\gamma_{m}^2}} \left[(4 M^2-Q^2) I_{-2}^B + 2M I_{-1}^B + I_0^B \right] \gamma_m\notag\\
&+ \frac{1}{\sqrt{1-\gamma_m^2}(r_+-r_-)} \sum_{\xi=\pm} \xi (2 M r_\xi-Q^2) \Bigg\{ a \gamma_m I_\xi^B \notag\\
&+ \left[ (a \gamma_m-\lambda_m) \left(M r_\xi-Q^2\right)
-\frac{r_{m5}^{(1)} r_{m6}^{(1)} }{2} [2M\gamma_m r_\xi-(a \lambda_m+\gamma_m Q^2)] \right] \left(\frac{I_\xi^B}{r_\xi^2}-\frac{I_{-1}^B}{r_\xi^2}-\frac{I_{-2}^B}{r_\xi} \right)\Bigg\} ,\label{I_t1}
\end{align}
and
\begin{align}
I_\phi^{B(0)}(r) =&\, \frac{\lambda_m}{\sqrt{1-\gamma_m^2}} I_0^B
+\frac{a}{\sqrt{1-\gamma_m^2}(r_+-r_-)}
\sum_{\xi=\pm} \xi \left[ 2M \gamma_m r_\xi-\left(a \lambda_m+\gamma_m Q^2 \right) \right]I_\xi^B ,\label{I_phi0}\\
I_\phi^{B(1)}(r) =&\, - \frac{1}{\sqrt{1-\gamma_m^2}} \left( \lambda_m \frac{r_{m5}^{(1)}r_{m6}^{(1)}}{2} I_{-2}^B + \gamma_m I_0^B \right)
+ \frac{a}{\sqrt{1-\gamma_m^2}(r_+-r_-)}
\sum_{\xi=\pm} \xi \Bigg\{ a \gamma_m I_\xi^B \notag\\
&+ \left[ (a \gamma_m-\lambda_m) \left(M r_\xi-Q^2\right)-\frac{r_{m5}^{(1)}r_{m6}^{(1)}}{2} \left(2 M \gamma_m r_\xi -a \lambda_m-\gamma_m Q^2 \right) \right] \left(\frac{I_{\xi}^B}{r_\xi^2}-\frac{I_{-1}^B}{r_\xi^2}-\frac{I_{-2}^B}{r_\xi}\right)\Bigg\} .\label{I_phi1}
\end{align}
Likewise,
\begin{align}
I_\psi^B(r) =&\, I_\psi^{B(0)}(r) +s_\parallel I_\psi^{B(1)}(r) ,\label{I_psi}
\end{align}
where
\begin{align}
I_\psi^{B(0)}(r) =&\, \frac{K\gamma_m-a (a \gamma_m-\lambda_m) }{\sqrt{1-\gamma_m^2}} \left( \frac{1}{\sqrt{K}} I_0^B
- {\rm Im}\left[I_{i+}^B \right] \right) ,\label{I_psi0}\\
I_\psi^{B(1)}(r) =&\, - \frac{r_{m5}^{(1)}r_{m6}^{(1)}}{2 K} \frac{ \left[K\gamma_m-a (a \gamma_m-\lambda_m) \right] }{\sqrt{1-\gamma_m^2}} {\rm Im}\left[I_{i+}^B \right] .\label{I_psi1}
\end{align}
The involved functions are
\begin{align}
I^B_{\pm}(r)&=\alpha^B \left[\frac{ F\left(\Upsilon^B_{\tau_m}|k^B \right) }{r_{m2}-r_{\pm}}+\frac{(r_{m2}-r_{m3}) \Pi\left(\beta^B_{\pm};\Upsilon^B_{\tau_m}\left|{k^{B}}\right)\right.}{(r_{m2}-r_{\pm})(r_{m3}-r_{\pm})}\right] -{\mathcal{I}^B_{\pm_i}} \label{IBpm}\;,\\
I^B_{i\pm}(r)& =\alpha^B \left[\frac{ F\left(\Upsilon^B_{\tau_m}|k^B \right) }{r_{m2}\mp i\sqrt{K}}+\frac{(r_{m2}-r_{m3}) \Pi\left(\beta^B_{i\pm};\Upsilon^B_{\tau_m}\left|{k^{B}}\right)\right.}{(r_{m2}\mp i\sqrt{K})(r_{m3}\mp i\sqrt{K})}\right] -{\mathcal{I}^B_{i\pm_i}} \label{Ii_pm_tau}\;,\\
I^B_{1}(r)&=\alpha^B\left[r_{m2} F\left(\Upsilon^B_{\tau_m}|k^B \right) +(r_{m3}-r_{m2})\Pi\left(\beta^B;\Upsilon^B_{\tau_m}\left|{k^{B}}\right)\right.\right] -{\mathcal{I}^B_{1_i}} \label{IB1}\;,\\
I^B_{2}(r)&= -\nu_{r} \frac{\sqrt{\left(r-r_{m1}\right)\left(r-r_{m2}\right)\left(r-r_{m3}\right)\left(r_{m4}-r\right)}}{r-r_{m2} } \nonumber\\
&-\frac{r_{m4}\left(r_{m3}-r_{m2}\right) -r_{m2}\left(r_{m3}+r_{m2}\right)}{\sqrt{(r_{m3}-r_{m1})(r_{m4}-r_{m2})}} F\left(\Upsilon^B_{\tau_m}|k^B \right) +\sqrt{(r_{m3}-r_{m1})(r_{m4}-r_{m2})} E\left(\Upsilon^B_{\tau_m}\left|{k^{B}}\right)\right. \nonumber\\
&+\frac{\left(r_{m3}-r_{m2}\right)\left(r_{m1}+r_{m2}+r_{m3}+r_{m4}\right)}{\sqrt{(r_{m3}-r_{m1})(r_{m4}-r_{m2})}} \Pi\left(\beta^B;\Upsilon^B_{\tau_m}\left|{k^{B}}\right)\right. -{\mathcal{I}^B_{2_i}} \label{IB2}\;
\end{align}
with incomplete elliptic integrals of the second kind $E(\varphi|k)$ and third kind $\Pi(n;\varphi|k)$.
The two additional integrals $I^B_{-1}$ and $I^B_{-2}$ are expressed by
\begin{align}
I^B_{-1}(r)&=\frac{\alpha^B}{r_{m2}}\left[ F\left(\Upsilon^B_{\tau_m}|k^B \right) -\frac{(r_{m3}-r_{m2})}{r_{m3}}\Pi\left(\beta^B_s;\Upsilon^B_{\tau_m}\left|{k^{B}}\right)\right.\right] -{\mathcal{I}^B_{-1_i}} \label{IB-1}\;,\\
I^B_{-2}(r)&=\frac{1}{r_{m3}} \Bigg[ \nu_{r} \frac{\sqrt{\left(r-r_{m1}\right)\left(r-r_{m2}\right)\left(r-r_{m3}\right)\left(r_{m4}-r\right)}}{r_{m1}r_{m4} \left(r_{m2}-r\right) r }\notag\\
&+\frac{r_{m4}\left(r_{m3}+r_{m2}\right)+r_{m2}\left(r_{m3}-r_{m2}\right)}{r_{m2}^2r_{m4}\sqrt{(r_{m3}-r_{m1})(r_{m4}-r_{m2})}} F\left(\Upsilon^B_{\tau_m}|k^B \right) +\frac{\sqrt{(r_{m3}-r_{m1})(r_{m4}-r_{m2})}}{r_{m1}r_{m2}r_{m4}}E\left(\Upsilon^B_{\tau_m}\left|{k^{B}}\right)\right. \nonumber\\
&+\frac{\left(r_{m2}-r_{m3}\right)}{r_{m2}\sqrt{(r_{m3}-r_{m1})(r_{m4}-r_{m2})}} \left(\frac{1}{r_{m1}}+\frac{1}{r_{m2}}+\frac{1}{r_{m3}}+\frac{1}{r_{m4}}\right) \Pi\left(\beta^B_s;\Upsilon^B_{\tau_m}\left|{k^{B}}\right)\right.\Bigg] -{\mathcal{I}^B_{-2_i}}. \label{IB-2}\;
\end{align}
The parameters of the elliptic functions above are given by
\begin{align}
\Upsilon_{\tau_m}^{B} &= {\rm am}\left(X^B(\tau_m) \mid k^{B}\right)
= \nu_{r_{i}}\sin^{-1}\left(\sqrt{\frac{(r-r_{m3})(r_{m4}-r_{m2})}{(r-r_{m2})(r_{m4}-r_{m3})}}\right),\\
\nu_{r}&={\rm sign}\Big(\frac{d r}{d\tau_m}\Big),\\
\beta_{\pm}^{B} &= \frac{(r_{m2}-r_{\pm})(r_{m4}-r_{m3})}{(r_{m3}-r_{\pm})(r_{m4}-r_{m2})},
\hspace*{12mm}
\beta_{i\pm}^{B}=\frac{(r_{m2}\mp i\sqrt{K})(r_{m4}-r_{m3})}{(r_{m3}\mp i\sqrt{K})(r_{m4}-r_{m2})},\\
\beta^{B} &= \frac{r_{m4}-r_{m3}}{r_{m4}-r_{m2}},
\hspace*{35mm}
\beta^{B}_s=\frac{r_{m2}}{r_{m3}}\frac{r_{m4}-r_{m3}}{r_{m4}-r_{m2}},\\
\alpha^B &= \frac{2}{\sqrt{(r_{m3}-r_{m1})(r_{m4}-r_{m2})}}. \label{Upsilon_m_b}
\end{align}
Notice that $\mathcal{I}^B_{0_i}$, $\mathcal{I}^B_{\pm_i}$, $\mathcal{I}^B_{i\pm_i}$, $\mathcal{I}^B_{1_i}$, $\mathcal{I}^B_{2_i}$, $\mathcal{I}^B_{-1_i}$, and $\mathcal{I}^B_{-2_i}$ depending on the initial conditions at $\tau_m = \tau_{mi} =0$ and $r=r_i$ are obtained by requiring ${I}_{0}^{B}(r_i)$, ${I}_{\pm}^{B}(r_i)$, ${I}_{i\pm}^{B}(r_i)$, ${I}_{1}^{B}(r_i)$, ${I}_{2}^{B}(r_i)$, ${I}_{-1}^{B}(r_i)$, and ${I}_{-2}^{B}(r_i)$ to be zero.
In the limit of $s_\parallel \rightarrow 0$ and $a\rightarrow 0$, the solutions reduce to those of \cite{wang-2022} and \cite{ciou-2025}, respectively. 
Moreover, they are finite and show no apparent singular integrals as $r\rightarrow r_{m3}$ or $ r\rightarrow r_{m4}$. 
This can be compared with the solutions in \cite{piovano-2025B}.

Following the discussion in \cite{ciou-2025}, the motion of the particle starts from $r_i=r_{m3}$ at $\tau_m(r_{m3})=0$, moves to $r=r_{m4}$ at $\tau_m(r_{m4})$, and then returns to $r=r_{m3}$.
The period of the full run with respect to Mino time $\tau_m$ is given by (\ref{I_r^B}), and can be written as
\begin{equation}
{\cal{T}}^B_r= 2 \tau_m(r_{m4})\equiv {\cal{T}}^{B(0)}_r + s_{\parallel} {\cal{T}}^{B(1)}_r \,, \label{periodR}
\end{equation}
where
\begin{align}
{\cal{T}}^{B(0)}_r =&\; \frac{4 K(k^B)}{\sqrt{(1-\gamma_m^2)(r_{m4}-r_{m2})(r_{m3}-r_{m1})}} , \\
{\cal{T}}^{B(1)}_r =& - \frac{r_{m5}^{(1)}r_{m6}^{(1)}}{r_{m3}\sqrt{1-\gamma_m^2}} \Bigg[ \frac{ r_{m4}\left(r_{m3}+r_{m2}\right)+r_{m2}\left(r_{m3}-r_{m2}\right)}{r_{m2}^2r_{m4}\sqrt{(r_{m3}-r_{m1})(r_{m4}-r_{m2})}} K\left(k^B \right) \notag\\
& +\frac{\sqrt{(r_{m3}-r_{m1})(r_{m4}-r_{m2})}}{r_{m1}r_{m2}r_{m4}} E\left({k^{B}}\right) \nonumber\\
& +\frac{\left(r_{m2}-r_{m3}\right)}{r_{m2}\sqrt{(r_{m3}-r_{m1})(r_{m4}-r_{m2})}} \left(\frac{1}{r_{m1}}+\frac{1}{r_{m2}}+\frac{1}{r_{m3}}+\frac{1}{r_{m4}}\right) \Pi\left(\beta^B_s \left|{k^{B}}\right)\right.\Bigg]
\end{align}
using the complete elliptic integrals:
\begin{equation} \label{complete_e_integral}
\begin{alignedat}{2}
F&\left(\Upsilon_{\tau_m}^B | k^B \right)\Big|_{r=r_{m4}} = K(k^B),
&\qquad
F&\left(\Upsilon_{\tau_m}^B | k^B \right)\Big|_{r=r_{m3}} = 0, \\
E&\left(\Upsilon_{\tau_m}^B | k^B \right)\Big|_{r=r_{m4}} = E(k^B),
&\qquad
E&\left(\Upsilon_{\tau_m}^B | k^B \right)\Big|_{r=r_{m3}} = 0, \\
\Pi&\left(\beta^B;\Upsilon_{\tau_m}^B | k^B \right)\Big|_{r=r_{m4}} = \Pi\left(\beta^B | k^B\right),
&\qquad
\Pi&\left(\beta^B;\Upsilon_{\tau_m}^B | k^B \right)\Big|_{r=r_{m3}} = 0 .
\end{alignedat}
\end{equation}
Then, we can find the corresponding radial frequency $\omega_r^B =2\pi /{\cal{T}}^B_r$ in Mino time.
Notice that the spin corrections to the period and frequency can also be obtained from the phase shift in (\ref{X1_tau}).
In addition, the periods about the other orbital variables are denoted as ${\cal{T}}_t^B =2\pi /\omega_t^B$, ${\cal{T}}_\phi^B =2\pi /\omega_\phi^B$, and ${\cal{T}}_{\delta \theta}^B =2\pi /\omega_{\delta \theta }^B$ for a full run, where their frequencies are defined by the Mino time average \cite{fujita-2009, van-de-meent-2020}:
\begin{align}
\omega_t^B &\equiv \left\langle \frac{dt}{d\tau_m} \right\rangle
= \frac{2}{{\cal{T}}^B_r} \int^{r_{m4}}_{r_{m3}} \frac{dt}{d\tau_m}\frac{dr'}{\sqrt{R_{m}^s(r')}} \,,\\
\omega_\phi^B &\equiv \left\langle \frac{d\phi}{d\tau_m} \right\rangle
= \frac{2}{{\cal{T}}^B_r} \int^{r_{m4}}_{r_{m3}} \frac{d\phi}{d\tau_m}\frac{dr'}{\sqrt{R_{m}^s(r')}} \,.
\end{align}
The induced polar motion along the variable $\theta^B ={\pi}/{2} +\delta\theta^B$ also has the spin precession frequency
\begin{align}
\omega_{\delta \theta}^B &\equiv \left\langle \frac{d\psi}{d\tau_m} \right\rangle
= \frac{2}{{\cal{T}}^B_r} \int^{r_{m4}}_{r_{m3}} \frac{d\psi}{d\tau_m}\frac{dr'}{\sqrt{R_{m}^s(r')}} \,,
\end{align}
obtained from (\ref{deltatheta2}).
Substituting the turning points $r_{m4}$ and $r_{m3}$ into (\ref{I_t0}), (\ref{I_t1}), (\ref{I_phi0}), and (\ref{I_phi1}), the frequencies are
\begin{align}
\omega_t^B &= \frac{2 I_t^{B(0)}(r_{m4})}{{\cal{T}}^{B(0)}_r} + s_\parallel \frac{2\left[ {\cal{T}}^{B(0)}_r I_t^{B(1)}(r_{m4}) - {\cal{T}}^{B(1)}_r I_t^{B(0)}(r_{m4}) \right]}{\left[{{\cal{T}}^{B(0)}_r} \right]^2} \label{omegaBt},\\
\omega_\phi^B &= \frac{2 I_\phi^{B(0)}(r_{m4})}{{\cal{T}}^{B(0)}_r} + s_\parallel \frac{2\left[ {\cal{T}}^{B(0)}_r I_\phi^{B(1)}(r_{m4}) - {\cal{T}}^{B(1)}_r I_\phi^{B(0)}(r_{m4}) \right]}{\left[{{\cal{T}}^{B(0)}_r} \right]^2} \label{omegaBphi}.
\end{align}
Their expressive forms can be written in terms of complete elliptic integrals (\ref{complete_e_integral}).
The first term with no explicit $s_\parallel$-dependence can be reproduced using Eqs. (40)-(43) in \cite{van-de-meent-2020} in the case of $\theta=\pi/2$ and $s_\parallel=0$.
From (\ref{I_psi0}) and (\ref{I_psi1}), the secondary spin precesses with a Mino time frequency,
\begin{align} \label{omegaBtheta}
\omega_{\delta \theta}^B = \frac{2 I_\psi^{B(0)}(r_{m4})}{{\cal{T}}^{B(0)}_r} + s_\parallel \frac{2\left[ {\cal{T}}^{B(0)}_r I_\psi^{B(1)}(r_{m4}) - {\cal{T}}^{B(1)}_r I_\psi^{B(0)}(r_{m4}) \right]}{\left[{{\cal{T}}^{B(0)}_r} \right]^2} .
\end{align}
With the above complete elliptic integrals (\ref{complete_e_integral}), when $\theta=\pi/2$ and $s_\parallel=0$, the first term can be identified as the frequency denoted by $\Upsilon_{\psi} = \Upsilon_{\psi,r} + \Upsilon_{\psi,z}$ shown in Eqs. (63)-(65) in \cite{van-de-meent-2020}. 

The frequencies of the orbits in the coordinate time $t^B$ are obtained below,
\begin{align} \label{Omega_i^B}
\Omega_r^B = \frac{\omega_r^B}{\omega_t^B}, \qquad
\Omega_\phi^B = \frac{\omega_\phi^B}{\omega_t^B}, \qquad
\Omega_{\delta \theta}^B = \frac{\omega_{\delta\theta}^B}{\omega_t^B}\, .
\end{align}
They serve as units for analyzing the gravitational-wave emission in the frequency domain. 
The corresponding periods can  be computed by $T_r^B =2\pi /\Omega_r^B$, $T_\phi^B =2\pi /\Omega_\phi^B$, and $T_{\delta \theta}^B =2\pi /\Omega_{\delta \theta}^B$.
Notice that reversing the orientation of the particle spin $s_\parallel$ changes the sign of the linear spin correction at the turning points $r_{m3}$, $r_{m4}$, and orbital frequencies above. However, as for $r_{m5}$ and $r_{m6}$, they  are a complex-conjugated pair, $r_{m5} = r^*_{m6}$, for $s_\parallel>0$ or  two real-valued roots,  $r_{m5} =- r_{m6}$, for $s_\parallel <0$.

\begin{figure}[h]
\centering
\includegraphics[width=13cm]{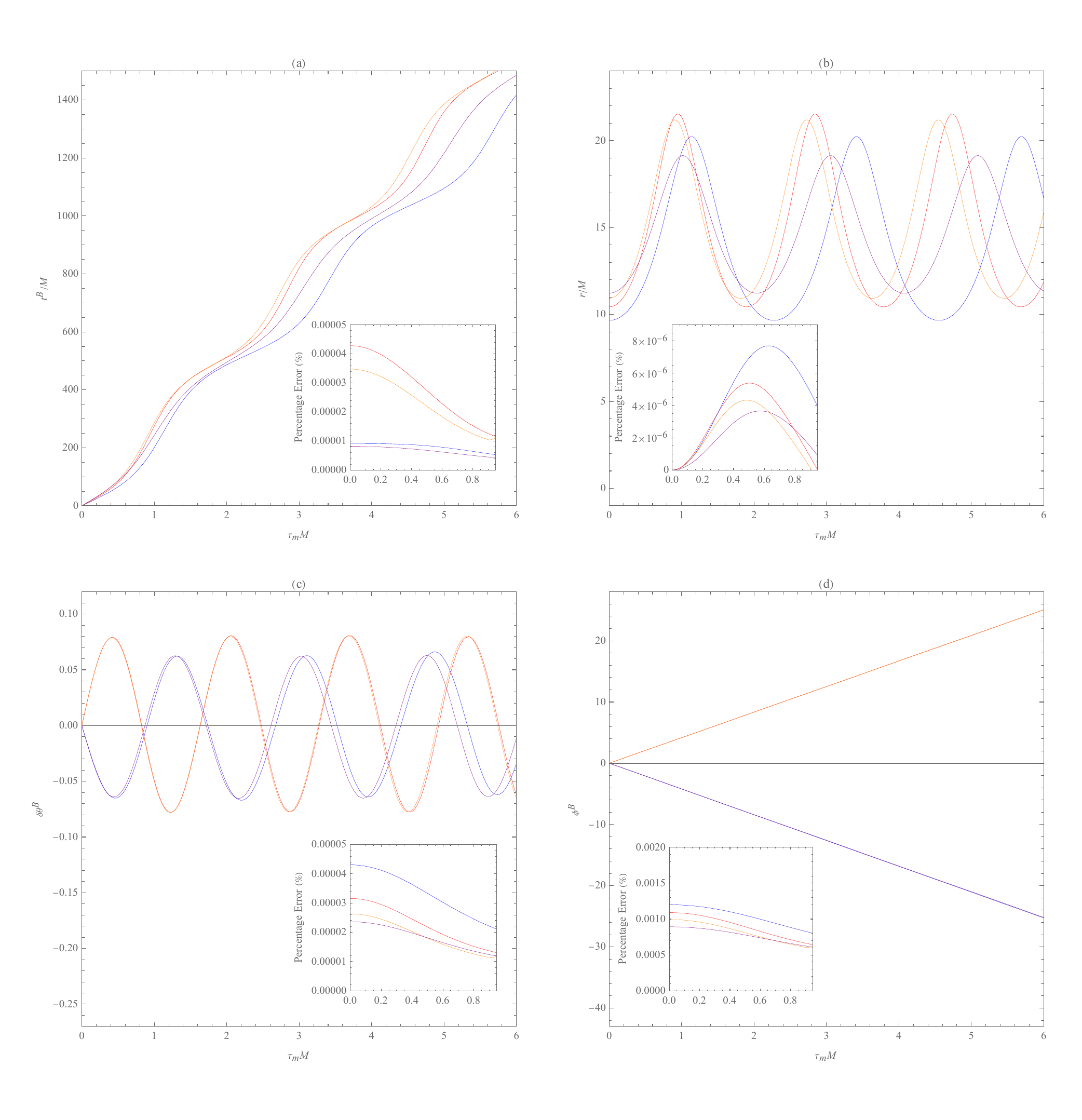}
    \caption{
    The plots show the dynamical variables of the orbits and corresponding percentage errors between analytical and numerical solutions: 
    (a) $t^B(\tau_m)$ from (\ref{I_t}) and (\ref{It});
    (b) $r(\tau_m)$ from (\ref{r_o}) and numerical inversion of (\ref{Ir});
    (c) $\delta \theta^B(\tau_m)$ from (\ref{deltatheta2}), (\ref{I_psi}), and (\ref{Ipsi});
    (d) $\phi^B(\tau_m)$ from (\ref{I_phi}) and (\ref{Iphi}) as a function of $\tau_m$ of the spinning particle traveling between two turning points for a misaligned spin, $s_\perp\neq 0$ with the initial conditions at $r_i=r_{m3}$ and $w=\pi/2$.
    The black hole parameters are chosen to be $a/M=0.5$, $Q/M=0.4$ ($a/M=0.5$, $Q/M=0.8$), and the particle parameters are chosen to be $\gamma_m=0.97$, $|\lambda_m|=0.42$, $s_\parallel/m= 0.1$, and $s/m=0.3$ for $\lambda_m>0$ (red) and $\lambda_m<0$ (blue) (for $\lambda_m>0$ (orange) and $\lambda_m<0$ (purple)).
    The error is defined by $|x -x_{\rm num}|/x_{\rm num} \times 100\% $, where $x=t$, $r$, $\psi$, $\phi$. }
\label{oscillation2}
\end{figure}

\begin{figure}[h]
\centering
\includegraphics[width=14cm]{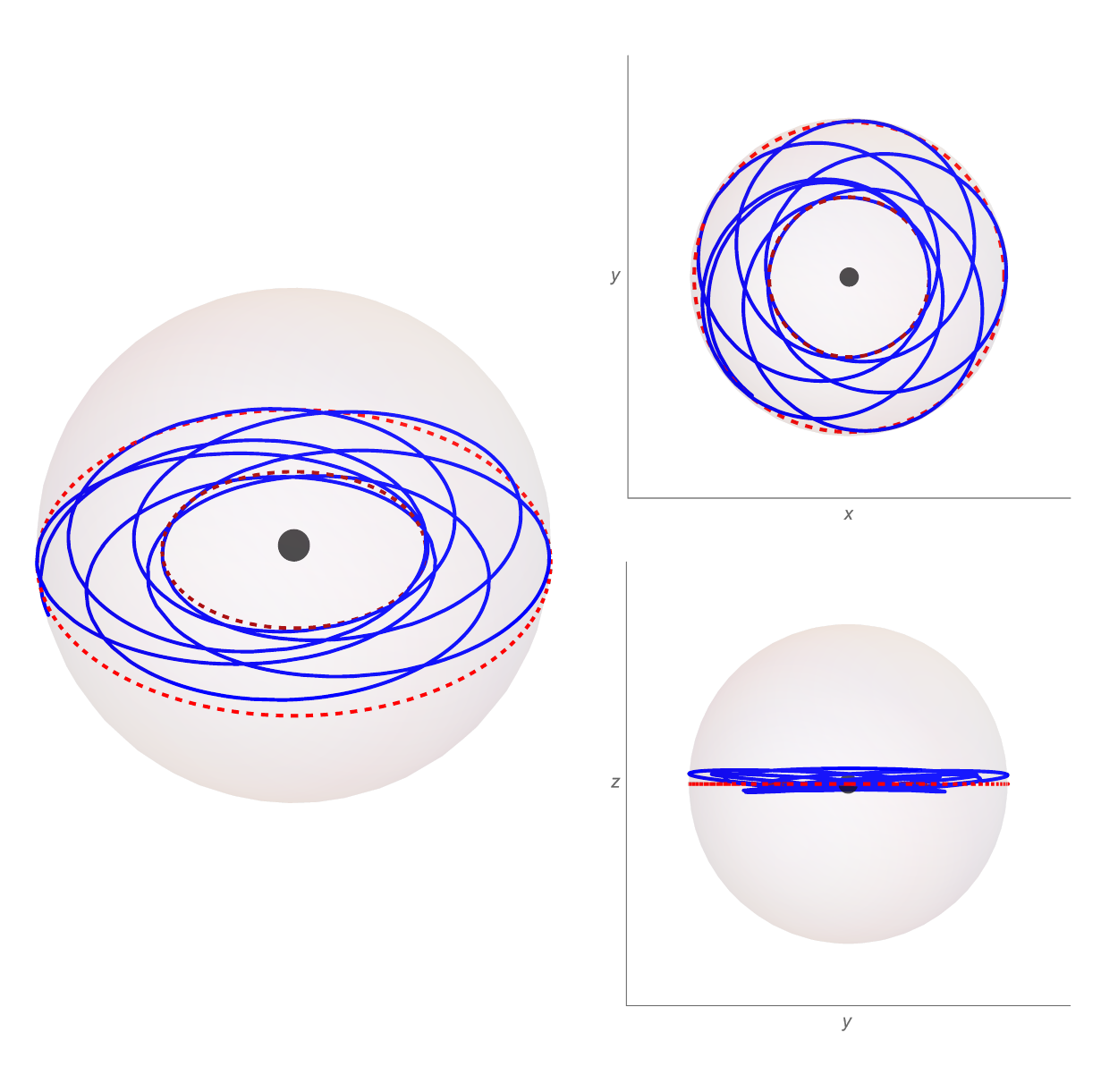}
    \caption{
    The plot shows the 3D trajectory of the particle traveling between two turning points, $r_{m4}$ (the red dashed circle) and $r_{m3}$ (the dark red dashed circle), with $s_\parallel/m = 0$, $s/m = 0.3$ around the Kerr–Newman black hole with $a/M=0.5$ and $Q/M=0.8$. 
    The induced oscillatory motion around the variable $\delta \theta^B$ is visible due to the misaligned spin $s_\parallel \neq s$. }
\label{oscillation1}
\end{figure}

Figure \ref{oscillation2} shows the evolution of the individual orbital variables $t$, $r$, $\theta$, and $\phi$, respectively, given by the analytic solutions with the parameters ($\gamma_m, \lambda_m$) of point B in Fig. \ref{KN radial potential and parameter space}.
This figure directly illustrates the dynamics of a spinning particle in the Kerr–Newman spacetime.
To further validate our analytical solutions, which is corrected up to linear spin order, we compare the evolution of each variable to its exact numerical integration and show good agreement.
Although higher-order spin corrections are included in the approximation given by (\ref{R_split}), they are insignificant and do not affect the reliability of the analytical solution to linear spin order.
Our goal is to derive compact analytical expressions that accurately produce the orbital evolution to a given order of spin.
Our approach to expressing the full solution differs from that developed in \cite{piovano-2025A}, where special care must be taken to deal with potential integration divergences at turning points through the choice of gauge.
In the Kerr limit as $Q \to 0$, the solutions formulated in \cite{drummond-2022A, piovano-2025A, skoupy-2025A, skoupy-2025B} are based on a specific gauge and orbital parametrization.
Therefore, a term-by-term analytical comparison is not straightforward.
However, the good agreement between our analytical solutions and the numerical integrations of the Mathisson–Papapetrou equations confirms their accuracy. 
The 3D trajectory is shown in Fig. \ref{oscillation1}.
%


In Fig. \ref{oscillation2}, for the parameters $a/M =0.5$ and $Q/M=0.8$, the effective perihelion and aphelion are approximately by $r_{m3} = 10.9M$ and $r_{m4} = 21.2M$, respectively.
The period of the radial motion from $r_{m3}$ to $ r_{m4}$ and back to $r_{m3}$ can be estimated by (\ref{periodR}), giving ${\cal{T}}_r^B \approx 1.8 /M$ in Mino time, where the spin correction contributes $\left| s_\parallel {\cal{T}}_r^{B(1)} \right| \approx 10^{-4} /M$.
Using (\ref{omegaBphi}),  the period of the motion about the azimuthal angle $\phi$ gives ${\cal{T}}^B_\phi =1.5 /M$.
As for the induced polar motion, from (\ref{omegaBtheta}) the period is about ${\cal{T}}^B_{\delta \theta}= 1.6 /M$.
Let us now provide an order of magnitude estimate about the gravitational-wave emission due to EMRIs in next section.

\section{ Order of magnitude estimate on gravitational-wave emission}
\label{sec4}
This section focuses on the observational consequences of the particle's trajectory as derived in the previous section.
Our goal is not to construct precise EMRI waveforms, but rather to show how the analytical orbital solutions derived in this work can be used directly in waveform calculations.
Specifically, we will provide an overview of using analytical orbital solutions to construct gravitational-wave signals. 
This establishes a qualitative relationship between the gravitational waveform and the parameters of the spinning particle. 

Following \cite{babak-2007}, we treat the Boyer–Lindquist coordinates as fictitious spherical polar coordinates and project the particle’s orbit onto Cartesian coordinates in flat spacetime,
\begin{equation} \label{Cartesian}
\begin{split}
x_p &= r \sin{\theta^B} \cos{\phi^B},\\
y_p &= r \sin{\theta^B} \sin{\phi^B},\\
z_p &= r \cos{\theta^B},
\end{split}
\end{equation}
where the orbital variables of a spinning particle $(t^B,\, r,\, \theta^B,\, \phi^B)$ are given by (\ref{I_t}), (\ref{I_phi}), (\ref{I_psi}), and (\ref{r_o}).
We then apply a flat-space wave-generation formula.
The input to a wave-generation formula is approximated by mapping the trajectory in the Boyer–Lindquist coordinates onto the resulting flat spacetime trajectory.
Consider the weak-field situation, where the spacetime metric is given by $g_{\mu\nu} =\eta_{\mu\nu} + h_{\mu\nu}$, in which $\eta_{\mu\nu}$ is the Minkowski metric and $h_{\mu\nu}$ is the small perturbation.
The {\it trace-reversed} metric perturbation is defined as $\bar h^{\mu\nu}\equiv h^{\mu\nu} -(1/2)\eta^{\mu\nu} h$, with $h = \eta^{\mu\nu}h_{\mu\nu}$.
In the Lorentz gauge $\partial_\alpha \bar{h}^{\mu\alpha}=0$, the linearized Einstein field equation is
\begin{eqnarray}
\Box \bar{h}^{\mu \nu}&=& -16\pi { T}^{\mu \nu} \label{waveq},
\end{eqnarray}
in which $\Box$ denotes the usual flat space wave operator. The effective energy-momentum tensor ${ T}^{\mu\nu}$ defined in flat spacetime satisfies the conservation law
\begin{eqnarray}
\partial_\mu { T}^{\mu\nu} &=& 0. \label{emcons}
\end{eqnarray}
The energy-momentum tensor \cite{babak-2007, cui-2025, zi-2023, gong-2025} due to the particle's orbit is given by
\begin{equation}
T^{\mu\nu}(t, {\vec{x}}) = m \frac{d x_p^\mu}{d\tau_m} \frac{d x_p^\nu}{d\tau_m} \frac{d\tau_m}{dt_p} \, \delta^3(\vec{x}-\vec{x}_p) ,
\end{equation}
where $m$ is the particle's mass and $\vec{x}_p$ is the position of the particle in (\ref{Cartesian}).

Considering observing gravitational waves at large distances from the sources, the transverse and traceless parts of the spatial components of $\bar{h}^{jk}$ are involved.
The solution can easily be written in terms of the retarded Green's function. 
The multipole expansion can be used to derive the quadrupole formula for a source located at the center ($\vec x=0$)
\begin{eqnarray}
\bar{h}^{jk}(t,\vec{x}) =
\frac2{|\vec{x}|} \left[ \ddot{I}^{jk} \right]_{t'=t- |\vec{x}|}\
\label{quadrupole}
\end{eqnarray}
with
\begin{eqnarray}
I^{jk}(t)
&=& \int x'^{j}x'^{k} \, T^{00}(t',\vec{x}') \, d^3x'
\label{quadmoment}
\end{eqnarray}
and $|\vec{x}|$ is the distance between the source and the observer.
For the extension to the quadrupole-octupole formula, see \cite{press-1977, bekenstein-1973}.
Finally, the analytical expression of the waveform is obtained as
\begin{equation}
\bar h^{jk} (t,\vec{x})
= \frac{2m}{|\vec{x}|} \frac{d^{2}}{d t^{2}}
\left[ {x}^{j}_p {x}^{k}_p \left( \frac{d t'_p}{d \tau_m} \right) \right]_{t'_p = t- |\vec{x}|}
\label{pressexp}
\end{equation}
in the Cartesian coordinates, evaluated at the retarded time $t_{ret} = t'_p = t-\vert \vec{x}\vert$ with $t_p'(\tau_m)$ in (\ref{tm}) and (\ref{I_t}).
Under the transverse-traceless gauge (TT gauge), the waveform can be further expressed by two independent polarization states, denoted as plus ($h_+$) and cross ($h_{\times}$).
These two states characterize how spacetime stretches and squeezes as the wave passes. 

Let us consider that the motion along the $r$ and $\theta$ coordinates is relatively slow compared with the $\phi$ coordinate, where we can approximate $x_p$ by $ x_p \approx r \cos(\Omega_\phi^B t)$.
According to (\ref{pressexp}), the waveform amplitude follows  $h \propto \frac{d^2}{dt^2} \left(x_p^i x_p^j \right)$, giving $ h \propto \left({\Omega_\phi^B}\right)^2 r^{2}$.
For an observer far from the source, the trajectory in Fig. \ref{oscillation1} can be regarded as effectively circular, provided that $|\vec x|$ is much greater than $r_{m3}$ and $r_{m4}$.
Naively, applying Kepler's third law $\left({\Omega_\phi^B}\right)^2 r^3 = constant$, the motion at $r=r_{m3}$ (an effective perihelion) experiences a stronger gravitational pull and contributes more significantly to the gravitational-wave emission than the motion at the aphelion at $r_{m4}$. 
This gives the waveform amplitude $ h \propto 1/r_{m3}$. 
Therefore, the smaller the perihelion, the larger the waveform amplitude.

Let us now provide an order of magnitude estimate about the gravitational-wave emission due to EMRIs.
Consider a Kerr–Newman black hole with spin $a/M=0.5$ and charge $Q/M=0.8$, and the surrounding low-mass celestial body with energy $\gamma_m =E_m/m = 0.97$, azimuthal angular momentum $\lambda_m =L_m/m= 4.2$, spin $s/m = 0.3$, and $s_\parallel/m = 0.1$.
Then, let us also consider the supermassive black hole Sagittarius A* with mass $M=4.1\times 10^6 M_{\odot}$, where $M_{\odot}$ refers to the mass of the Sun.
Therefore, according to (\ref{periodR}), (\ref{omegaBt}), (\ref{omegaBphi}), (\ref{omegaBtheta}) and (\ref{Omega_i^B}), the period $T_r^B=487.3 M$ is about $9847.9$ (sec) and the frequency $\Omega_r^B$ is $0.0006$ (Hz).
Also, the period $T_\phi^B=403.4 M$ is about $8151.8$ (sec) and the frequency $\Omega_\phi^B$ is $0.0008$ (Hz), and the period $T_{\delta \theta}^B=438.9 M$ is about $8871.0$ (sec) and the frequency $\Omega_{\delta \theta}^B$ is $0.0007$ (Hz).
They play a key role in determining the typical frequency of the emitted gravitational waves, providing a reference frequency scale for  binary systems under observation. 
In particular, the LISA observation is within the frequency range of $10^{-4}$ (Hz) to $1$ (Hz) in \cite{danzmann-2000, prince-2003, amaro-seoane-2012}.
Moreover, the orbital frequencies can be separated into $\Omega_{i}^B = \Omega_{i}^{B(0)} +s_\parallel \Omega_{i}^{B(1)}$ for $i=r, \phi, \delta\theta$ by (\ref{Omega_i^B}). 
Using the representative parameters of Sgr A* and the spinning particle above, the particle-spin correction modifies the orbital frequencies by about
\begin{equation}
\begin{split}
&\frac{\left|\Omega_r^B - \Omega_r^{B(0)}\right|}{\Omega_r^B} \times 100\% \approx \frac{1.8 \times 10^{-7} \,\text{(Hz)}}{0.0006 \,\text{(Hz)}} \times 100\% \approx 0.03 \% \,, \\
&\frac{\left|\Omega_\phi^B - \Omega_\phi^{B(0)}\right|}{\Omega_\phi^B} \times 100\% \approx \frac{1.7 \times 10^{-5} \,\text{(Hz)}}{0.0008 \,\text{(Hz)}} \times 100\% \approx 2.3 \% \,, \\
&\frac{\left|\Omega_{\delta\theta}^B - \Omega_{\delta\theta}^{B(0)}\right|}{\Omega_{\delta\theta}^B} \times 100\% \approx \frac{2.6 \times 10^{-8} \,\text{(Hz)}}{0.0007 \,\text{(Hz)}} \times 100\% \approx 0.004 \% \,.
\end{split}
\end{equation}
These corrections are larger than the respective errors about various variables between our analytical solutions and the numerical integrations of the Mathisson–Papapetrou equations, as shown in Fig. \ref{oscillation2}. These are reliable results.
Although the corrections may be too small to distinguish, EMRIs observed by LISA accumulate orbital phase over multiple cycles.
Consequently, even tiny frequency shifts could lead to measurable phase differences, suggesting that the effects of particle spin could become relevant in the modelling of future high-precision waveforms. 
Recalling the quadrupole formula (\ref{pressexp}), the order of magnitude of the waveform amplitude can be
\begin{equation}
h \sim \frac{2}{|\vec{x}|} \left( \frac{Gm}{c^2} \right) \frac{G M}{c^2} \frac{1}{r_{m3}} ,
\end{equation}
or by Kepler's third law
\begin{equation}
h \sim \frac{2}{|\vec{x}|} \left( \frac{Gm}{c^2} \right) \left( \frac{G M}{c^2} \right)^{2/3}\left({\Omega_\phi^B}\right)^{2/3} ,
\end{equation}
where again $r_{m3}$ is an effective perihelion.
A distance $|\vec{x}| = 26000$ (ly) is adopted between the supermassive black hole Sgr A* and the observer.
Then, given the test particle's mass $m/M=10^{-6}$, the amplitude $h$ is about $3 \times 10^{-18}$, or the amplitude $h\, |\vec{x}|/m$ is $0.12$.\\

\section{concluding remarks}
\label{sec5}
In this paper, we extend the work of \cite{ciou-2025} to consider a spinning particle orbiting a Kerr–Newman black hole.
The dynamics is governed by the Mathisson–Papapetrou equations in the pole-dipole approximation, which includes spin-curvature coupling to the first order of spin.
In terms of conserved quantities, the dynamical equations in Mino time can be transformed into the integral forms for both aligned and misaligned spins with respect to the orbital motion.
The radial potential is derived to study the parameter space of the particle for various types of orbit, based on the roots with the corrections from the particle's spin.

Here, we focus primarily on the equatorial motion of a particle oscillating between two turning points, which are the two outermost roots of the radial potential, in the misaligned case.
In this case, there is an induced oscillatory motion, $\delta \theta$, out of the equatorial plane.
Two turning points can be regarded as the effective perihelion and aphelion of the orbits.
The orbital solutions can be written in terms of the elliptic integrals and the Jacobian elliptic functions of manifestly real functions in Mino time.
The periods and frequencies of motion in the $r$, $\delta \theta$, and $\phi$ directions are obtained, providing predictions of spin-induced shifts on the order of magnitude. 
Furthermore, when the orbit becomes a source of gravitational-wave emission, the periods of motion will provide essential input in determining the gravitational waves in the frequency domain.
An estimate of the amount of gravitational waves emitted by EMRIs is given.

These analytical expressions provide a useful starting point for future studies of gravitational waves emitted by EMRIs.
We plan to achieve this by including the radiation reaction and self-force effects. 
Future work will also extend the homoclinic and inspiral orbits in the Reissner–Nordstr\"om black hole \cite{ciou-2025} to the Kerr–Newman black hole.
In particular, numerical kludge waveforms due to the sources mentioned above will be constructed to study their significant properties, such as potential chaos.

\begin{acknowledgments}
This work was supported in part by the National Science and Technology Council (NSTC) of Taiwan, Republic of China.
\end{acknowledgments}

\appendix
\section{The roots of the radial potential $\mathcal{R}_m(r)$ }
\label{appendixA}
When $s_{\parallel} = 0$, we write the radial potential $\mathcal{R}_m (r)$ from  \cite{wang-2022} as follows,
\begin{align}\label{R_m}
\mathcal{R}_m({r})=S_m {r}^4+T_m {r}^3+U_m {r}^2+V_m {r}+W_m,
\end{align}
where
\begin{align}
&S_m=\gamma_{m}^2-1,\\
&T_m=2M,\\
&U_m=a^2(\gamma_{m}^2-1)-\lambda_{m}^2-Q^2,\\
&V_m=2M(a\gamma_{m}-\lambda_{m})^2,\\
&W_m=-Q^2(a\gamma_{m}-\lambda_{m})^2.
\end{align}
Furthermore, it is more transparent to rewrite the radial potential using its roots, namely
\begin{align}
\mathcal{R}_m({r}) =\left(\gamma_m^2-1\right) \left({r}-r_{m1}^{(0)}\right) \left({r}-r_{m2}^{(0)}\right) \left({r}-r_{m3}^{(0)}\right) \left({r}-r_{m4}^{(0)}\right)\, .
\end{align}
By the Ferrari's method, four roots of this quartic polynomial are expressed as
\begin{align}
r_{m1}^{(0)}&=-\frac{M}{2\left(\gamma_m^2-1\right)}+z_m+\sqrt{-\hspace*{1mm}\frac{\textbf{X}_m}{2}-z_m^2-\frac{\textbf{Y}_m}{4z_m}},\\
r_{m2}^{(0)}&=-\frac{M}{2\left(\gamma_m^2-1\right)}+z_m-\sqrt{-\hspace*{1mm}\frac{\textbf{X}_m}{2}-z_m^2-\frac{\textbf{Y}_m}{4z_m}},\\
r_{m3}^{(0)}&=-\frac{M}{2\left(\gamma_m^2-1\right)}-z_m+\sqrt{-\hspace*{1mm}\frac{\textbf{X}_m}{2}-z_m^2+\frac{\textbf{Y}_m}{4z_m}},\\
r_{m4}^{(0)}&=-\frac{M}{2\left(\gamma_m^2-1\right)}-z_m-\sqrt{-\hspace*{1mm}\frac{\textbf{X}_m}{2}-z_m^2+\frac{\textbf{Y}_m}{4z_m}},
\end{align}
where
\begin{align}
z_m=\sqrt{\frac{\Omega_{m +}+\Omega_{m -}-\frac{\textbf{X}_m}{3}}{2}}\ ,
\end{align}
with
\begin{align}
\Omega_{m \pm}=\sqrt[3]{-\hspace*{1mm}\frac{\textbf{Q}_m}{2}\pm\sqrt{\left(\frac{\textbf{P}_m}{3}\right)^3+\left(\frac{\textbf{Q}_m}{2}\right)^2}}\ ,
\end{align}
and
\begin{align}
\textbf{P}_m&=-\hspace*{1mm}\frac{\textbf{X}_m^2}{12}-\textbf{Z}_m,\\
\textbf{Q}_m&=-\hspace*{1mm}\frac{\textbf{X}_m}{3}\left[\left(\frac{\textbf{X}_m}{6}\right)^2-\textbf{Z}_m\right]-\hspace*{1mm}\frac{\textbf{Y}_m^2}{8}.
\end{align}
In addition, $X_m$, $Y_m$, and $Z_m$ are the short notations for
\begin{align}
\textbf{X}_m&=\frac{8U_m S_m -3T_m^2}{8S_m^2},\\
\textbf{Y}_m&=\frac{T_m^3-4U_m T_m S_m+8V_m S_m^2}{8S_m^3},\\
\textbf{Z}_m&=\frac{-3T_m^4+256W_m S_m^3-64V_m T_m S_m^2+16U_m T_m^2S_m}{256S_m^4}.
\end{align}
The sum of the roots satisfies the relation
\begin{equation} \label{root_relation}
r_{m1}^{(0)} +r_{m2}^{(0)} +r_{m3}^{(0)} +r_{m4}^{(0)} =-\frac{2M}{\gamma_m^2-1}.
\end{equation}

\bibliography{References}

\end{document}